\documentclass{aastex}

\usepackage{amsmath}
\usepackage{amsfonts}
\usepackage{bm}	
\usepackage{lscape}
\usepackage{color}
\usepackage{lscape}
\def\arcsec{^{\prime\prime}}
\def\arcmin{^{\prime}}
\def\farcs{\hbox{$.\!\!^{\prime\prime}$}}

\def\gtrsim{\mathrel{\hbox{\rlap{\hbox{\lower4pt\hbox{$\sim$}}}\hbox{$>$}}}}
\def\lessim{\mathrel{\hbox{\rlap{\hbox{\lower4pt\hbox{$\sim$}}}\hbox{$<$}}}}
\def\JB{{\rm Jy\,beam^{-1}}}
\def\mJB{{\rm mJy\,beam^{-1}}}
\def\kms{{\rm km\ s^{-1}}}
\def\Ms{M_{\odot}}
\def\II{{\rm I\hspace{-0.1em}I}}

\slugcomment{}

\shorttitle{Infall and Keplerian motion in TMC-1A} 
\shortauthors{Aso et al.}

\begin{document}


\title{ALMA Observations of the Transition from Infall Motion to Keplerian Rotation around the Late-phase Protostar TMC-1A}
\author{Yusuke Aso\altaffilmark{1}, Nagayoshi Ohashi\altaffilmark{2,3}, Kazuya Saigo\altaffilmark{4}, Shin Koyamatsu\altaffilmark{1}, Yuri Aikawa\altaffilmark{5}, Masahiko Hayashi\altaffilmark{6}, Masahiro N. Machida\altaffilmark{7}, Masao Saito\altaffilmark{8,9}, Shigehisa Takakuwa\altaffilmark{3}, Kengo Tomida\altaffilmark{10}, Kohji Tomisaka \altaffilmark{6}, and Hsi-Wei Yen\altaffilmark{3}}

\altaffiltext{1}{Department of Astronomy, Graduate School of Science, The University of Tokyo, 731 Hongo, Bunkyo-ku, Tokyo 113-0033, Japan}
\altaffiltext{2}{Subaru Telescope, National Astronomical Observatory of Japan, 650 North A'ohoku Place, Hilo, HI 96720, USA}
\altaffiltext{3}{Academia Sinica Institute of Astronomy and Astrophysics, P.O. Box 23-141, Taipei 10617, Taiwan}
\altaffiltext{4}{Chile Observatory, National Astronomical Obervatory of Japan, Osawa 2-21-1, Mitaka, Tokyo 181-8588, Japan}
\altaffiltext{5}{Department of Earth and Planetary Sciences, Kobe University, Kobe 657-8501, Japan}
\altaffiltext{6}{National Astronomical Observatory of Japan, Osawa, 2-21-1, Mitaka, Tokyo 181-8588, Japan}
\altaffiltext{7}{Department of Earth and Planetary Sciences, Faculty of Sciences Kyushu University, Fukuoka 812-8581, Japan}
\altaffiltext{8}{Nobeyama Radio Observatory, Nobeyama, Minamimaki, Minamisaku, Nagano 384-1305, Japan}
\altaffiltext{9}{SOKENDAI, Department of Astronomical Science, Graduate University for Advanced Studies}
\altaffiltext{10}{Department of Astrophysical Sciences, Princeton University, Princeton, NJ 08544, USA}

\begin{abstract}
We have observed the Class I protostar TMC-1A with Atacama Millimeter/submillimeter Array (ALMA) in the emissions of $^{12}$CO and C$^{18}$O $(J=2-1)$, and 1.3-mm dust continuum. Continuum emission with a deconvolve size of $0\farcs50\times 0\farcs37$, perpendicular to the $^{12}$CO outflow, is detected. It most likely traces a circumstellar disk around TMC-1A, as previously reported. In contrast, the C$^{18}$O a more extended structure is detected in C$^{18}$O although it is still elongated with a deconvolved size of $3\farcs3\times 2\farcs2$, indicating that C$^{18}$O traces mainly a flattened envelope surrounding the disk and the central protostar. C$^{18}$O shows a clear velocity gradient perpendicular to the outflow at higher velocities, indicative of rotation, while an additional velocity gradient along the outflow is found at lower velocities. The radial profile of the rotational velocity is analyzed in detail, finding that it is given as a power-law $\propto r^{-a}$ with an index of $\sim 0.5$ at higher velocities. This indicates that the rotation at higher velocities can be explained as Keplerian rotation orbiting a protostar with a dynamical mass of $0.68\ \Ms$ (inclination corrected). The additional velocity gradient of C$^{18}$O along the outflow is considered to be mainly infall motions in the envelope. Position-Velocity diagrams made from models consisting of an infalling envelope and a Keplerian disk are compared with the observations, revealing that the observed infall velocity is $\sim$0.3 times smaller than free fall velocity yielded by the dynamical mass of the protostar. Magnetic fields could be responsible for the slow infall velocity. A possible scenario of Keplerian disk formation is discussed.
\end{abstract}

\section{Introduction}
Protoplanetary disks play key roles in the evolution of young stellar objects (YSOs). 
In the early stages of star formation, mass and angular momentum are transferred to the central stars through protoplanetary disks.
In the later stages, these disks can become the sites of planet formation \citep[e.g.][]{ho2013}.
Protoplanetary disks are ubiquitous around classical T-Tauri stars (CTTSs) or Class $\II$ sources and have been observed in thermal dust continuum and molecular line emissions at millimeter and submillimeter wavelengths since the 1990s.
Keplerian rotation is revealed for many of the disks \citep{gu.du1998,gu1999,si2000,qi2004,hu2009,ro2010}.
In the evolutionary stages earlier than T-Tauri stars, accreting protostars (Class I or 0 sources) are associated with disk-like structures that exhibit continuum and line emissions.
These disk-like structures often show infall motions with little rotation \citep[e.g.][]{jo2009,en2009,en2011,br2009,ye2010,ye2011} and are called ``pseudo disks'' (Galli and Shu, 1993ab), although pseudo disks have not been firmly identified with observations. Recent high-sensitivity interferometric observations, however, have also revealed Keplerian disks around Class 0 and Class I protostars \citep{ta2012,ye2013,ha2014,cho2010,to2012}, that are still deeply embedded in protostellar envelopes.

Despite the ubiquity of Keplerian disks in various protostellar stages, their formation process from protostellar envelopes is still controversial both theoretically and observationally. 
Theoretical calculations of collapsing isothermal spheres with ideal, non-magnetized conditions, such as ``inside-out collapse'' models \citep{te1984,sh1987}, provide a conventional picture of star and disk formation. According to these models, a central protostar forms with a Keplerian disk at the center of a slowly rotating dense molecular cloud core as the gas begins dynamical infall, which conserves its angular momentum. 
Larger angular momentum is gradually transferred from the outer region of the dense core and is re-distributed in the disk as time goes on. 
There are some observational studies confirming this tendency that a Keplerian disk develops later so that the rotational velocity and size of the disk become larger in later evolutionary stages of YSOs \citep[e.g.][]{oh1997,ye2013}.
Although these studies observationally suggest the growth of Keplerian disks in the course of YSO evolution, the dynamics of each source are not spatially resolved in these studies. This prevent us from directly investigating how a disk develops in a protostellar system. 
Recent theoretical simulations for disk formation around protostars, including MHD calculations, are still under heated discussion. It is reported that such simulations have difficulties in transforming envelopes to Keplerian disks because of strong magnetic braking or fragmentation 
\citep{me.li2008,me.li2009}. However, \citet{tom2013,to2015,ma2011} point out that non-ideal MHD effects can suppress angular momentum transport by magnetic fields and enable formation of circumstellar disks in the early phase of star formation \citep[see also][]{jo2012}.
The formation of Keplerian disks is closely related to the dynamical transition from the infall motion in protostellar envelopes to the Keplerian rotation. In fact, such a transition is being observationally revealed around several protostars by the most recent studies \citep{mu2013,ch2014,oh2014,ta2014,ye2014}. The number of samples is, however, still {\rm too} limited to perform any statistical study. In order for us to unveil the disk formation mechanism, it is essential to observe the two kinds of dynamics (infall and rotation) around protostars and increase the number of direct detections of the transition between the two.

In order to investigate the expected transition from the infall motion to Keplerian rotation, we made observations of the Class I protostar TMC-1A (IRAS 04365+2535) with the Atacama Large Millimeter/Submillimeter Array (ALMA) in $^{12}$CO $J=2-1$ (230.5380 GHz), C$^{18}$O $J=2-1$ (219.5604 GHz), and 1.3-mm continuum emission. As its name suggests, our target TMC-1A is in the Taurus molecular cloud, which is one of the closest star-forming regions ($d=140$ pc). Its bolometric luminosity $L_{\rm bol}=2.7\ L_{\odot}$ and bolometric temperature $T_{\rm bol}=118$ K \citep{kr2012} indicate that TMC-1A is typical Class I protostar. The systemic velocity of TMC-1A was measured at $V_{\rm LSR}=6.4\ \kms$ by \citet{te1989}. 

A molecular outflow ($t_{\rm dyn}\sim 2.5\times 10^{3}\ {\rm yr}$) is associated with TMC-1A \citep{hi1995, ta1996, ch1996}.
Interferometric observations detected a dense gas condensation toward the protostar with its velocity gradient perpendicular to the outflow axis \citep{oh1996, oh1997}. Furthermore Submillimeter Array (SMA) observations of this protostellar envelope show that higher-velocity components ($|V|\gtrsim 1.4\ \kms$) are located closer to the central protostar than the lower-velocity components ($|V|\lessim 0.8\ \kms$), suggesting a differential rotation of the envelope \citep{ye2013}. 
Their analysis, using the Position-Velocity (PV) diagram with a cut perpendicular to the outflow axis to obtain information purely about rotation by avoiding contamination from outflowing or infalling gas, showed that TMC-1A has a shallower rotational velocity profile ($p=-0.6\pm 0.1$ with $v\propto r^{p}$) and higher rotation velocity ($1.0-4.0\ \kms$) than their other samples such as L1527 IRS and L1448-mm on a 1000 AU scale. 
They suggest that TMC-1A should have a Keplerian disk ($p=-0.5$) with $M_{*}=1.1\pm 0.1\ \Ms$ and a large angular momentum should have been transferred into the disk from the surrounding envelope. 
This result can be understood in terms of the conventional analytical model with inside-out collapse. Based on this model, the rotation profile indicates that TMC-1A is in an evolutionary stage at $t\sim 5\times 10^{5}$ yr after the onset of collapse. 
A similar analysis to \citet{ye2013} was applied to the visibilities of TMC-1A data obtained with Plateau de Bure Interferometer \citep{ha2014}. 
They measured gas kinematics in uv domain and suggested that the rotational velocity profile in the inner part $(r<80-100\ {\rm AU})$ is explained with the power-law index of $p=-0.5$ (i.e. Keplerian disk) while the outer part is explained with $p=-1$. 
They also estimated the size of the disk by removing a spherical-infalling envelope component from the continuum emission toward TMC-1A, although it was not well constrained ($R_{\rm disk}=80-220$ AU). 
These pieces of previous work indicate that TMC-1A is in the disk-developing phase, and hence is a good target to investigate the formation process of Keplerian disks around protostars.

Our observations and data reduction are described in Sec.\ref{sc:obs}. In Sec.\ref{sc:rslt}, we present the continuum and molecular-line results. In Sec.\ref{sc:ana}, we analyze velocity structures of the C$^{18}$O line, in particular rotating motions, while we discuss infalling motions around TMC-1A in more detail in Sec.\ref{sc:dsc}. In Sec.\ref{sc:cnc}, we present a summary of the results and our interpretation. In addition, an investigation into the analysis using a PV diagram and a comparison of flattened and spherical envelope models are presented in Appendix \ref{sc:app1} and \ref{sc:app3}.

\section{ALMA Observations}
\label{sc:obs}
We observed our target, TMC-1A during Cycle 0 using the Atacama Large Millimeter/Submillimeter Array (ALMA) on November 6, 2012. The configuration is ``extended'' with 23 antennas. Extended configuration covers projected uv distances from 16 m to 285 m (11.7$-$208.7 k$\lambda$ in C$^{18}$O $J=2-1$). This minimum baseline resolves out more than 50\% of the flux if a structure is extended more than $7\farcs7$ \citep{wi.we1994}, corresponding to $\sim 1100$ AU at the distance to TMC-1A. The coordinate of map center is $\alpha$(J2000)=4$^{\rm h}$39$^{\rm m}$35$^{\rm s}\!\!$.010, $\delta$(J2000)=25$^{\circ }41\arcmin 45\farcs 500$. $^{12}$CO $(J=2-1)$ and C$^{18}$O $(J=2-1)$ lines and 1.3-mm continuum emission in Band 6 were observed for 63 minutes (on source). In order to derive high velocity resolutions for molecular line observations, we configured the correlator in Frequency Division Mode (FDM) for four spectral windows. Each spectral window has 3,840 channels covering 234 MHz bandwidth. Emission-free channels are used to make the continuum map centered at 225 GHz. The total integrated frequency width of the continuum map is $\sim 920$ MHz. The observed visibilities were Fourier transformed and CLEANed with Common Astronomy Software Applications (CASA). In this process, we adopted Briggs weighting with the robust parameter of 0.5 and binned 2 channels thus the frequency (velocity) resolution in this paper is 122 kHz corresponding to 0.16 $\kms$ in $^{12}$CO ($J=2-1$) and 0.17 $\kms$ in C$^{18}$O ($J=2-1$). We also set CLEAN boxes to enclose only positive emission in dirty maps. The synthesized beam sizes of the CLEANed maps are $1\farcs 02\times 0\farcs 90$ for the $^{12}$CO line, $1\farcs 06\times 0\farcs 90$ for the C$^{18}$O line, and $1\farcs 01\times 0\farcs 87$ for the continuum. J0522-364, J0510+180, and Callisto were observed as Passband, Gain, and Flux calibrator. In regard to the flux calibrator Callisto, models of solar system objects were updated from Butler-Horizons-2010 to Butler-JPL-2012 for CASA 4.0 after the delivery of our data and the fractional difference (2012/2010 $-1$) was reported. Following the value (Callisto: $-0.15$, 241 GHz), we multiplied the delivered intensities by 0.85. The rms noise levels are measured in channels where emission is detected to take the noise due to spatial filtering artifacts as well as the thermal noise into account. The parameters of our observations mentioned above (and others) are summarized in Table \ref{tb:1}.

\section{Results}
\label{sc:rslt}
\subsection{Continuum and $^{12}$CO}
\label{sc:cnco}
\begin{figure}
\epsscale{1.0}
\plotone{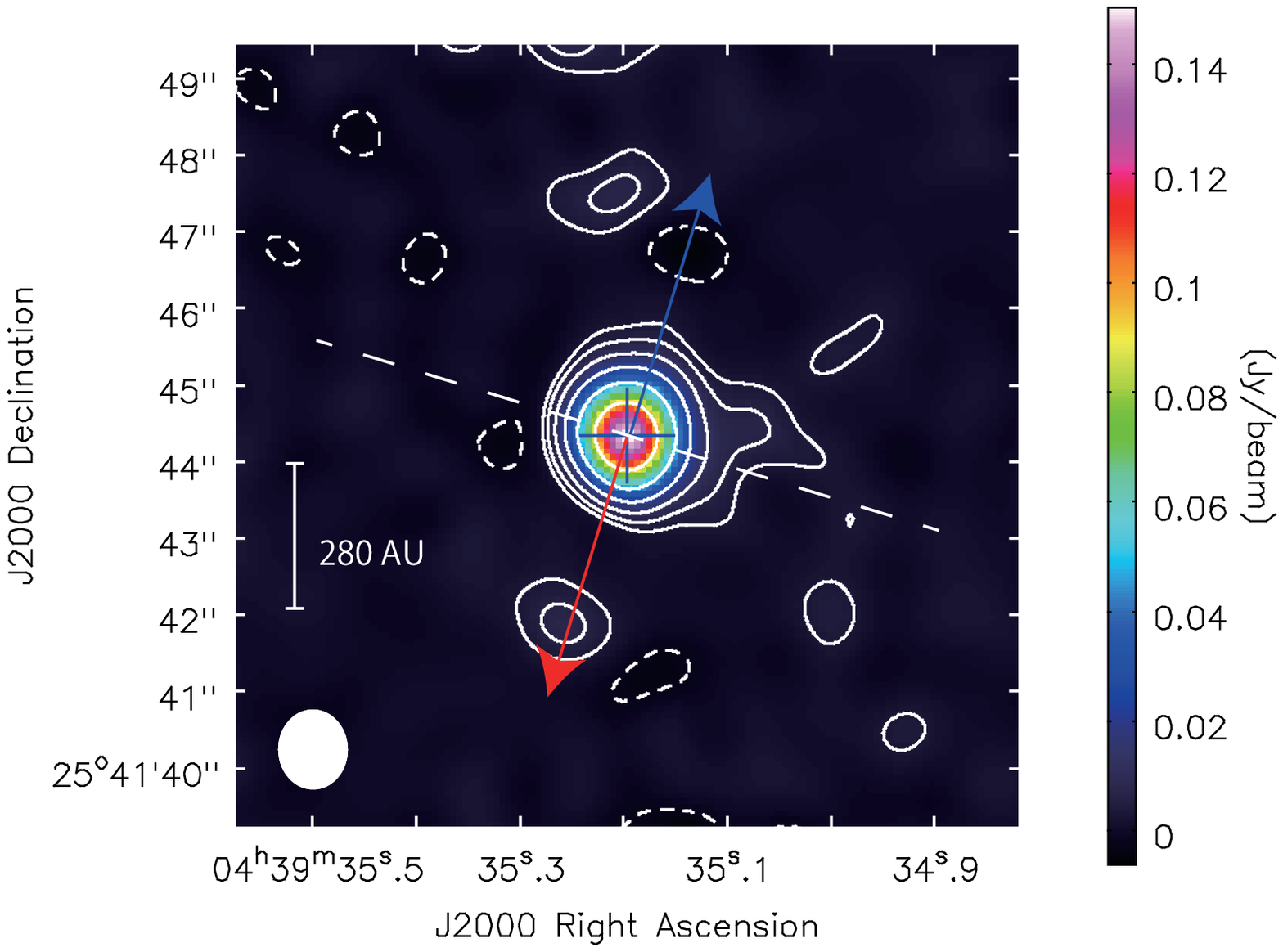}
\caption{Continuum emission map of TMC-1A observed with ALMA. Contour levels are $-3,3,6,12,24, ... \times \sigma$ where 1$\sigma$ corresponds to $0.96\ \mJB$. A cross shows the position of the continuum emission peak. A filled ellipse at the bottom left corner denotes the ALMA synthesized beam; $1\farcs 01\times 0\farcs 87,\ {\rm P.A.}=+0.87^{\circ}$. The deconvolved size of the continuum emission is $0\farcs 50 \pm 0\farcs03\times 0\farcs 37\pm0\farcs05,\ {\rm P.A}=+73^{\circ}\pm 16^{\circ}$ and the elongation direction ($73^{\circ}$) is shown with a white dashed line. Blue and red arrows show the direction of the molecular outflow from TMC-1A observed in $^{12}$CO $(J=2-1)$ line (Fig. \ref{fig:12}).\label{fig:con}}
\end{figure}

Fig. \ref{fig:con} shows the 225 GHz continuum emission in TMC-1A observed with ALMA. Strong continuum emission with a weak extension to the west is detected. Though the size of the continuum emission is almost similar to the beam size ($1\farcs 01\times 0\farcs 87$), its deconvolved size can be estimated from a 2D Gaussian fitting to be $0\farcs 50\pm 0\farcs 03\times 0\farcs37\pm 0\farcs 05$, P.A.$=73^{\circ}\pm 16^{\circ}$. The peak position is also measured from the Gaussian fitting to be $\alpha$(2000)=$4^{\rm h}39^{\rm m}35.20^{\rm s}$, $\delta$(2000)=$+25^{\circ}41\arcmin 44\farcs 35$, which is consistent with previous measurements (Yen et al. 2013). The measured continuum peak position is used as the central protostar position of TMC-1A in this paper. The peak intensity and the total flux density of the continuum emission derived from a Gaussian fitting are $148.5\pm 2.2\ \mJB$ and $182.1\pm2.6\ {\rm mJy}$ respectively. The dust mass calculated with the total flux density is $M_{\rm dust}=(4.2\pm 0.9)\times 10^{-4}\ \Ms $ assuming a standard opacity coefficient $\kappa (1\ {\rm THz})=10.0\ {\rm cm}^{2}\,{\rm g}^{-1}$ 
\citep{be1990}, an opacity index $\beta =1.46$, and a dust temperature $T_{c}=28$ K \citep{ch1998}, which is a temperature estimated from a fitting to the SED of TMC-1A at $\lambda \gtrsim 20\ \mu {\rm m}$. Thus, the gas mass is estimated to be $4.2\times 10^{-2}\ \Ms$ if assuming the gas/dust mass ratio is 100.

\begin{figure}
\epsscale{1.0}
\plotone{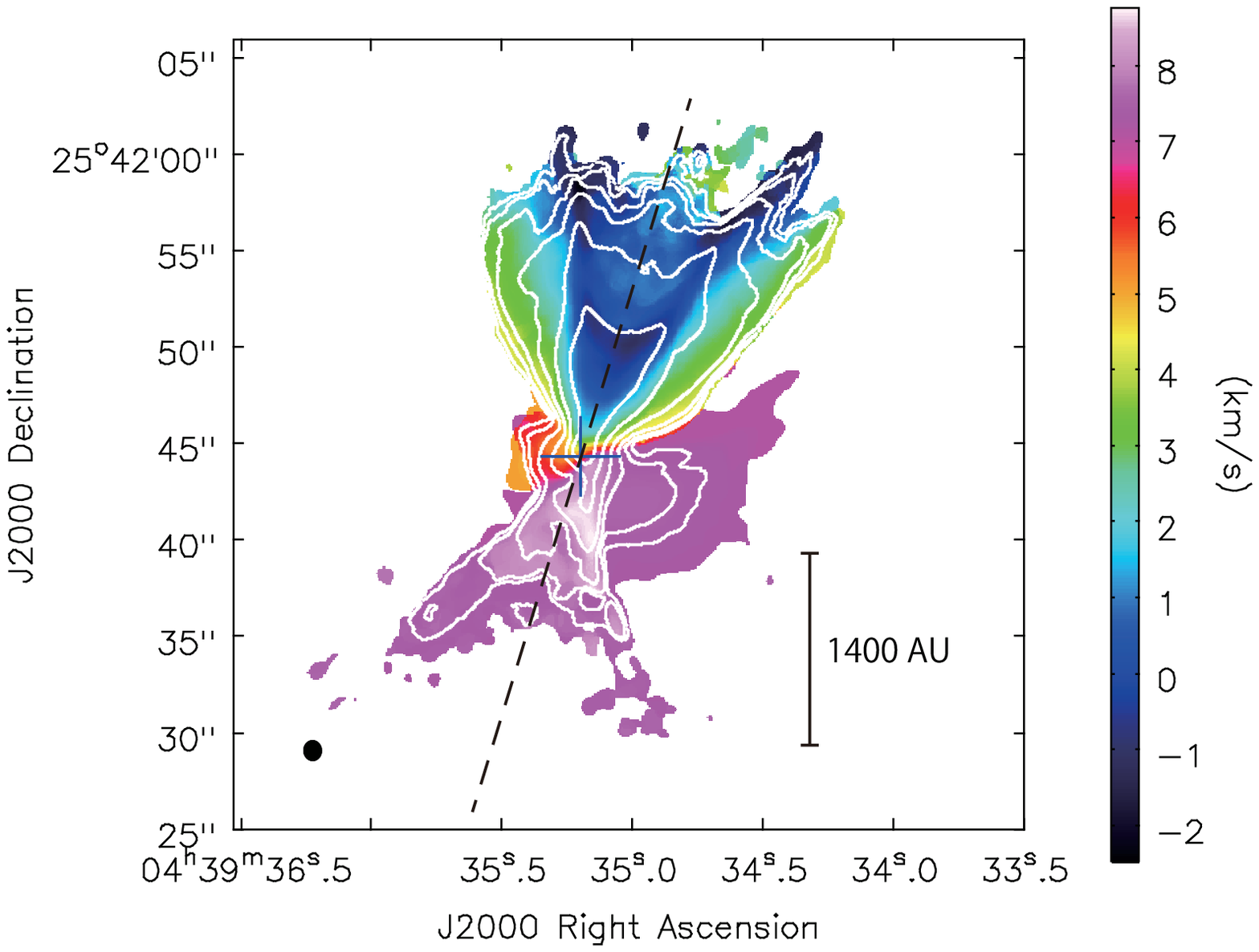}
\caption{Moment 0 ($white\ contours$) and 1 ($color$) maps of the $^{12}$CO $(J=2-1)$ emission in TMC-1A. Contour levels of the moment 0 map are $-3,3,6,12,24, ... \times \sigma$ where 1$\sigma$ corresponds to $0.034\ \JB \, \kms$. A Cross shows the position of the central protostar (continuum emission peak). A filled ellipse at the bottom-left corner denotes the ALMA synthesized beam; $1\farcs 02\times 0\farcs 90,\ {\rm P.A.}=-178^{\circ}$. $^{12}$CO clearly traces the molecular outflow from TMC-1A and the axis of the outflow ($\sim -17^{\circ}$) is shown with a dashed line.\label{fig:12}}
\end{figure}

\begin{figure}
\epsscale{1.0}
\plotone{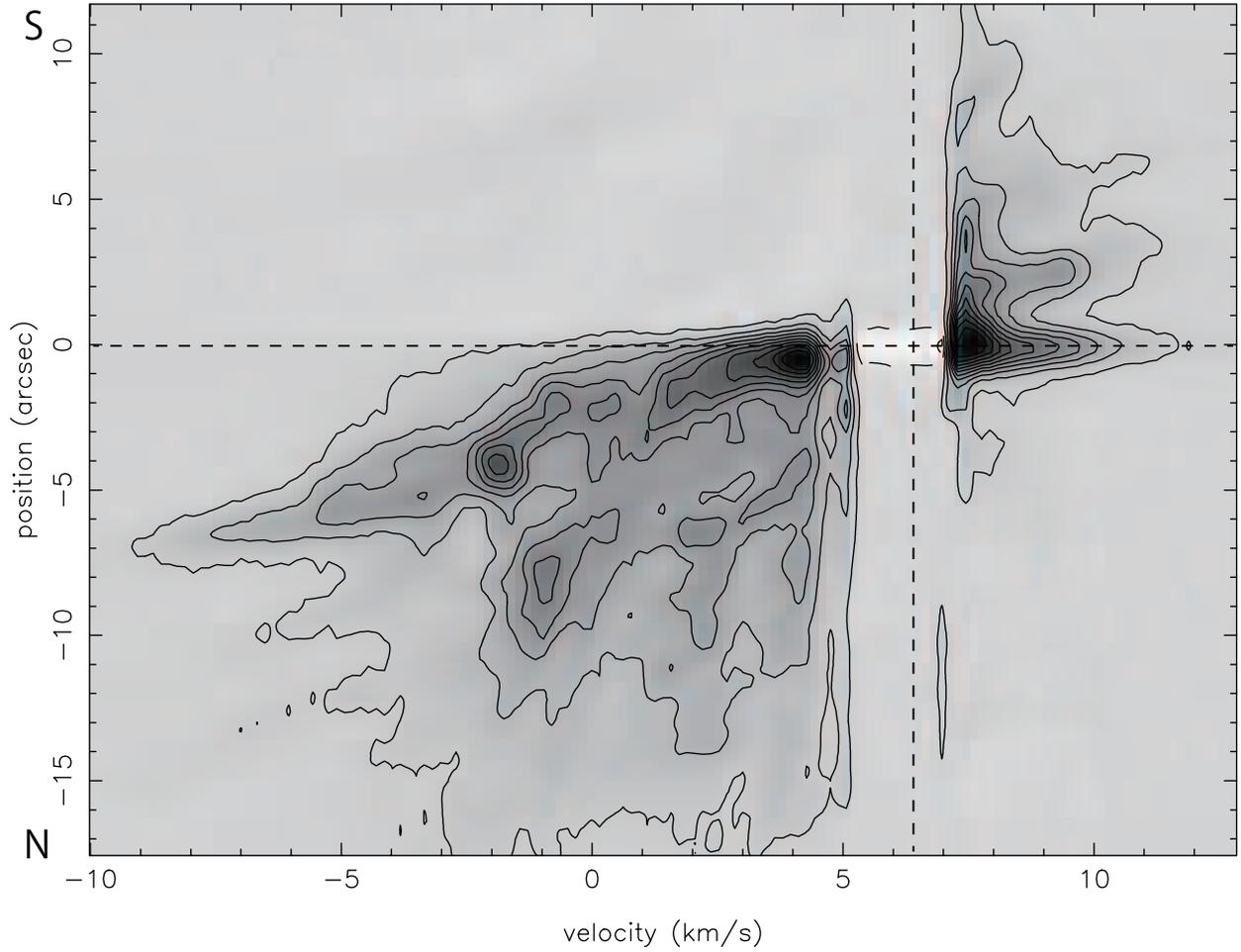}
\caption{Position-Velocity diagrams of the $^{12}$CO $(J=2-1)$ emission in TMC-1A along the outflow axes (PA $=-17^{\circ}$). Contour levels are $6\sigma $ spacing from $3\sigma $ ($1\sigma =20\ \mJB$). Central vertical dashed lines show the systemic velocity ($V_{\rm sys}=6.4\ \kms$) and central horizontal dashed lines show the central position. \label{fig:12pv}}
\end{figure}

Fig. \ref{fig:12} shows the integrated intensity (moment 0) and the intensity-weighted mean velocity (moment 1) maps of the $^{12}$CO $(J=2-1)$ emission toward TMC-1A, presented respectively in contour and in color. The $^{12}$CO emission is detected at more than 3$\sigma$ level at the velocity ranging from $-9\ \kms$ to $12\ \kms$ in Local Standard of Rest frame (LSR), except for the velocity from $5\ \kms$ to $7\ \kms$, where the $^{12}$CO emission is most likely resolved out. This resolved-out velocity range is around the systemic velocity $V_{\rm sys}=6.4\ \kms$. The emission shows a bipolar structure and a shell-like shape. The overall shape and velocity gradient seen in the maps indicate that the $^{12}$CO emission clearly traces a molecular outflow going in the north-south direction. The position angle of the outflow axis is measured to be $-17^{\circ}$ by eye based on the direction of the velocity gradient and the shape of the $^{12}$CO emission, as indicated with the dashed line in Fig. \ref{fig:12}, which is consistent with previous papers \citep[e.g. $-10^{\circ}$;][]{ch1996} and perpendicular to the elongation direction of the continuum emission. The blue lobe is detected more clearly than the red lobe. As suggested by the previous $^{12}$CO ($J=1-0$) observation (Tamura et al. 1996), an accelerated motion along the outflow axis can be seen, especially in the blue lobe on $\sim1000$ AU scale as shown in Fig. \ref{fig:12pv} where the Position-Velocity diagram of the $^{12}$CO emission along the outflow axis is presented. Essentially, the velocity increases out along the outflow axis (``Hubble law'') from the central protostar. 
On the other hand, across the outflow axis, the higher-velocity component is located closer to the outflow axis. This feature across the outflow axis has been observed in other outflows as well, and can be understood by an outflow driven by a parabolic wide-angle wind \citep[e.g.][]{le2000}. In addition to the $^{12}$CO component tracing the outflow, there seems to be another $^{12}$CO component near the central star, which shows an elongated structure almost perpendicular to the outflow axis. This additional component also has a velocity gradient along its extension. The geometrical and kinematical structures of this additional $^{12}$CO component are similar to those of C$^{18}$O $(J=2-1)$, as will be shown later.

The compactness of the continuum emission with its position angle perpendicular to the outflow axis suggests that the continuum emission arises from a compact disk and the major axis of the expected disk should be at P.A.$=73^{\circ}$ direction (white dashed line in Fig. \ref{fig:con}). Although the uncertainty of the elongation direction of the continuum emission is relatively large ($\pm16^{\circ}$), we hereafter adopt this position angle as the disk major axis around TMC-1A. 
The ratio of minor axis to major axis corresponds to the inclination angle of $i=42^{\circ}$ and $55^{\circ}$ if assuming a geometrically thin disk and $H/R=0.2$, respectively.

\subsection{C$^{18}$O $J=2-1$}
\label{sc:18}
\begin{figure}
\epsscale{1.0}
\plotone{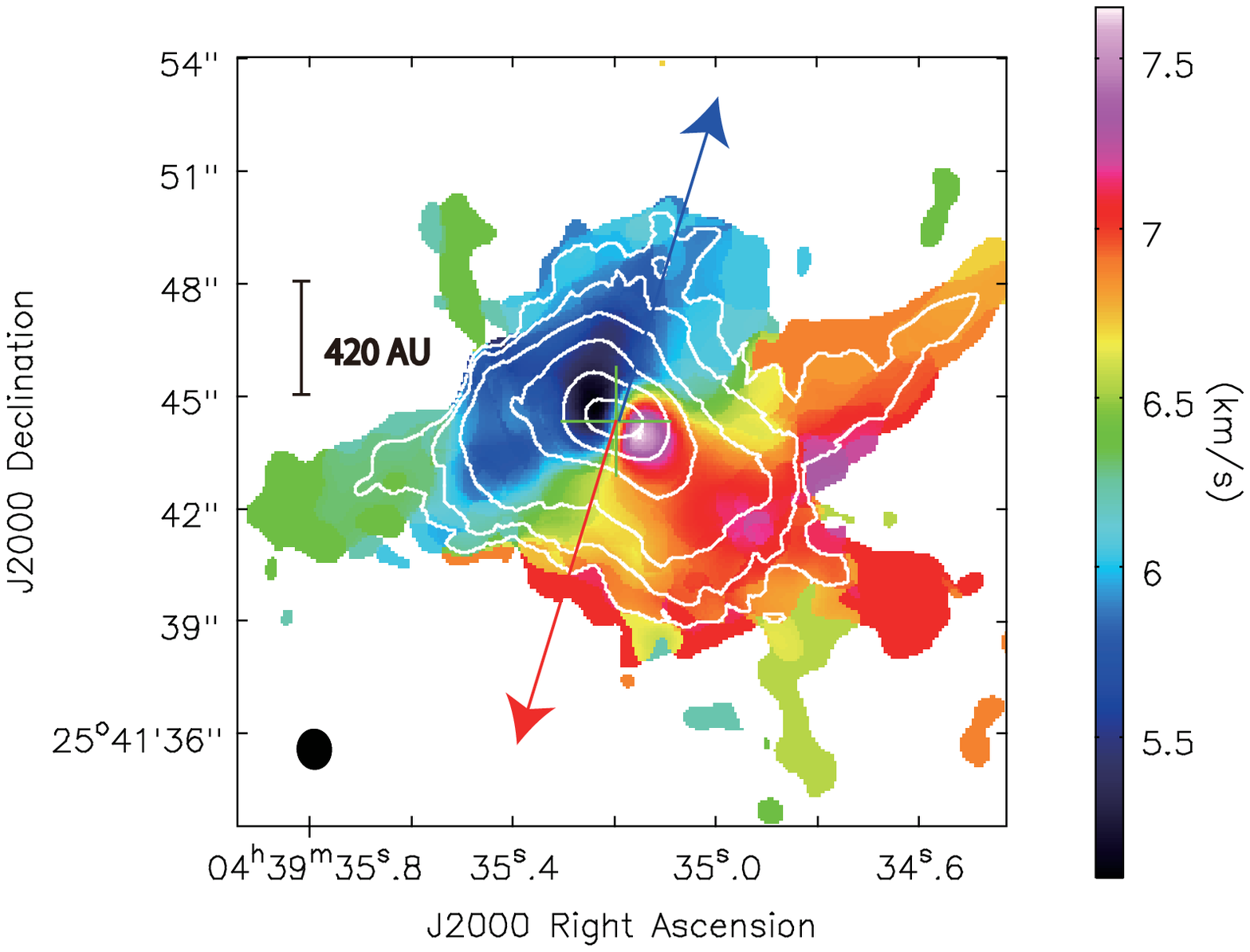}
\caption{Moment 0 ($white\ contours$) and 1 ($color$) maps of the C$^{18}$O $(J=2-1)$ emission in TMC-1A. Contour levels of the moment 0 map are $-3,3,6,12,24, ... \times \sigma$ where 1$\sigma$ corresponds to $8.1\ \mJB \, \kms$. A cross shows the position of the central protostar (continuum emission peak). A filled ellipse at the bottom-left corner denotes the ALMA synthesized beam; $1\farcs 06\times 0\farcs 90,\ {\rm P.A.}=-176^{\circ}$. Blue and red arrows show the direction of the molecular outflow from TMC-1A observed in $^{12}$CO $(J=2-1)$ line (Fig. \ref{fig:12}).\label{fig:mom}}
\end{figure}

C$^{18}$O $(J=2-1)$ is detected at more than a 3$\sigma $ level at LSR velocities ranging from $2.7\ \kms$ to $10.4\ \kms$. In Fig. \ref{fig:mom}, the moment 0 (MOM0) map integrated over this velocity range is shown in white contours, overlaid on the moment 1 (MOM1) map shown in color. The overall distribution of the C$^{18}$O integrated intensity exhibits an elongated structure almost perpendicular to the molecular outflow, with a peak located at the protostar position. The deconvolved size of MOM0 map is $3.3\arcsec \pm 0.1 \arcsec \times 2.2\arcsec \pm 0.1 \arcsec$ with P.A.$=67^{\circ}\pm 2^{\circ}$. This morphology showing elongation perpendicular to the outflow axis (P.A.=$-17^{\circ}$) indicates that the C$^{18}$O emission mainly traces a flattened envelope around TMC-1A, as was also suggested by \citet{ye2013}. In addition, there are weak extensions to the north, the northwest, and the east. The overall velocity gradient is seen from northeast to southwest, which is almost perpendicular to the outflow axis. These results are quite consistent with previous observations in C$^{18}$O $(J=2-1)$ using SMA (Yen et al. 2013) although their map did not show the weak extensions detected in our ALMA observations. 

\begin{figure}
\epsscale{1.0}
\plotone{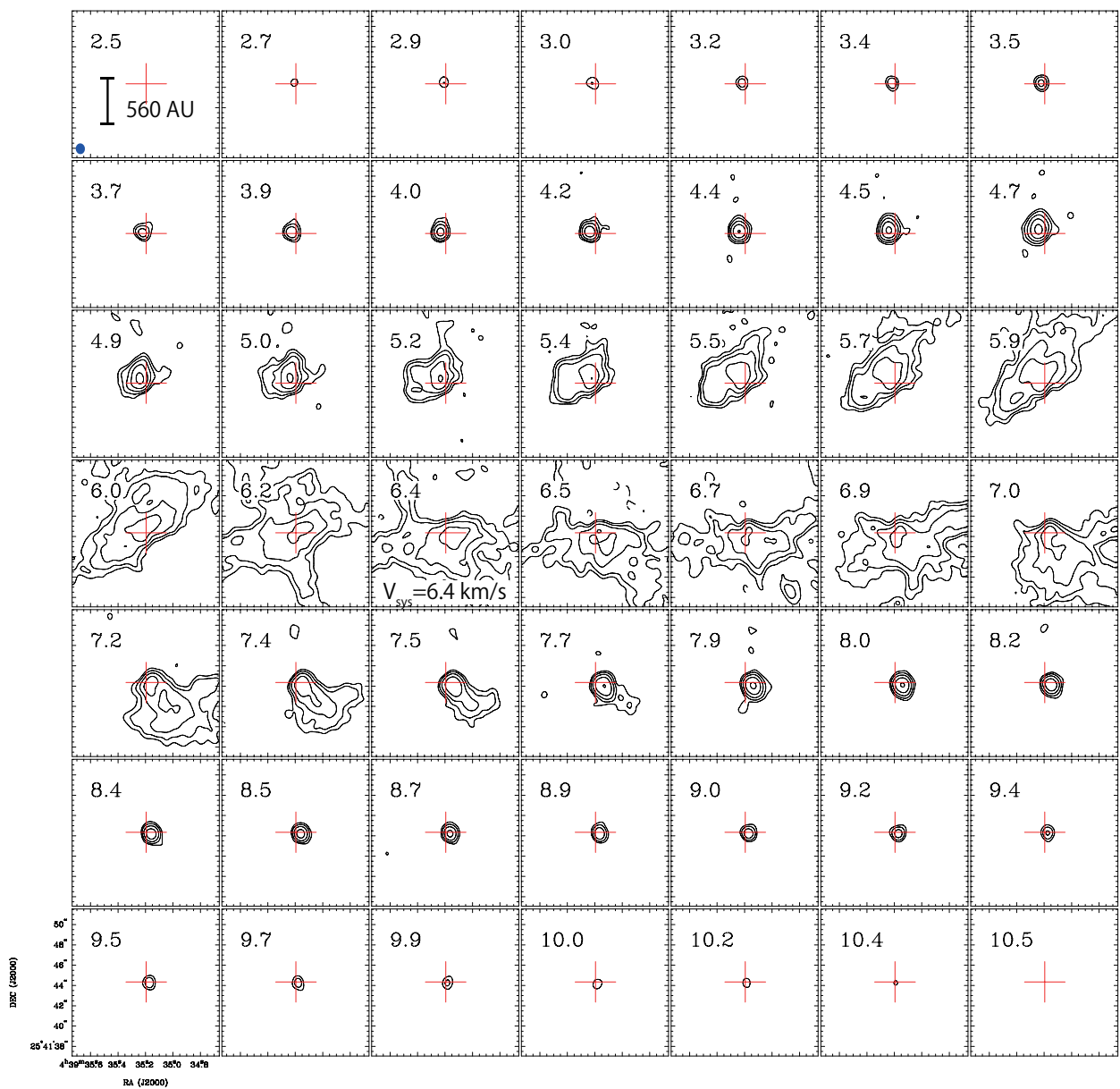}
\caption{Channel maps of the C$^{18}$O $(J=2-1)$ emission in TMC-1A. Contour levels are $-3,3,6,12,24, ... \times \sigma$ where 1$\sigma$ corresponds to $7.1\ \mJB$. Crosses show the position of the central protostar (continuum emission peak). A filled ellipse in the top left panel denotes the ALMA synthesized beam; $1\farcs 06\times 0\farcs 90,\ {\rm P.A.}=-176^{\circ}$. LSR velocities are shown at the top-left corner of each panel and the systemic velocity is 6.4 $\kms$.\label{fig:ch}}
\end{figure}

In order for us to see more detailed velocity structures, channel maps shown in Fig. \ref{fig:ch} are inspected. At high velocities, $V_{\rm LSR}\leq 4.9\ \kms$ and $\geq 7.9\ \kms$ (which are blueshifted and redshifted respectively by more than $1.5\ \kms$ with respect to the systemic velocity of $6.4\ \kms$), there are compact emission with strong peaks located nearby the protostar position. The sizes of these emission are smaller than $\sim 3\arcsec$ or 420 AU. The peaks of these emission are located on the east side of the protostar at blueshifted velocities, while those for redshifted velocities are located on the west side, making a velocity gradient from east to west. This direction is roughly perpendicular to the outflow axis (P.A.$=-17^{\circ}$). On the other hand, at lower blueshifted and redshifted velocities ($5.0\ \kms \leq V_{\rm LSR}\leq 6.0\ \kms$), extended structures become more dominant. At lower blueshifted velocities, structures elongated from northwest to southeast with additional extensions to the northeastern side and the southern side appear, while at lower redshifted velocities, structures elongating to southwest become more dominant. At velocities close to the systemic velocity, weak emission appears in even more extended structures to east, south, west, and north, which seem to form an X-shaped structure as a whole. A similar X-shaped structure can be also seen in MOM0 and MOM1 maps although they are less obvious because of a dilution effect after the channel integration.

\begin{figure}
\epsscale{1.0}
\plotone{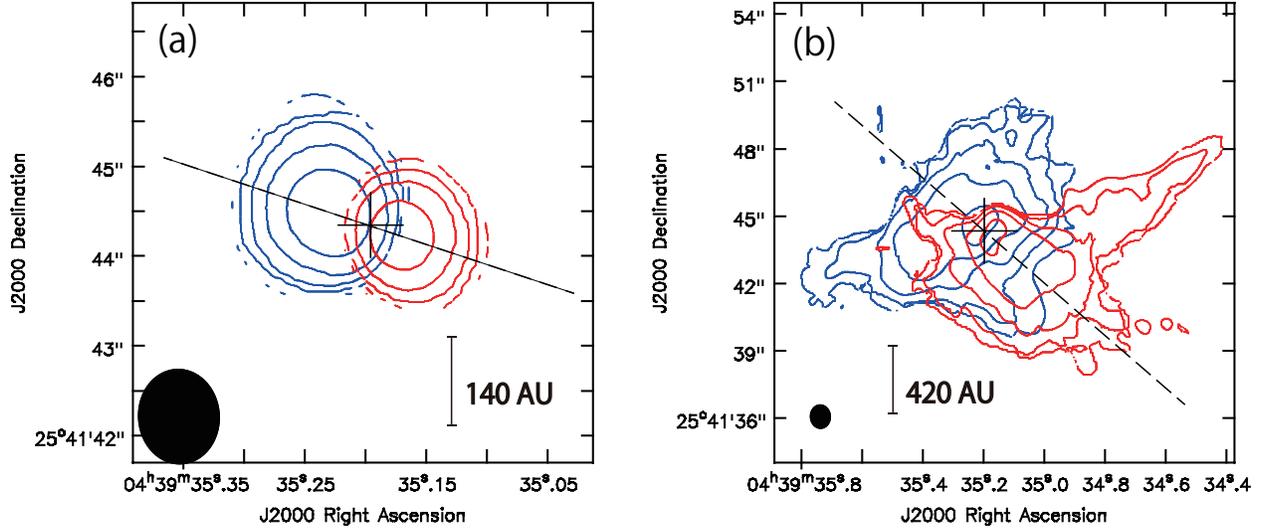}
\caption{Maps of blueshifted (blue contours) and redshifted (red contours) emission of (a) high-velocity ($|\Delta V|>2.0\ \kms$) and (b) low-velocity ($|\Delta V|<1.0\ \kms$) components of C$^{18}$O $(J=2-1)$ emission in TMC-1A as observed with ALMA. Contour levels are $-2,2,4,8,16,32 \times \sigma$ where 1$\sigma$ corresponds to $2.9\ \mJB$ and $4.1\ \mJB$ for (a) and (b), respectively. Crosses in both panels show the position of the central protostar (continuum emission peak). Filled ellipses at the bottom left corner in both panels are the same as Fig. \ref{fig:mom}. The dashed line passing peaks of blueshifted and redshifted maps in each panel shows the direction of the velocity gradient.\label{fig:ce}}
\end{figure}

Since the C$^{18}$O emission shows both compact emission at higher velocities and extended emission at lower velocities (whose origin and nature may be different from each other), integrated channel maps of blueshifted and redshifted emission, as presented in Fig. \ref{fig:ce}, were made for further investigation: Fig. \ref{fig:ce}a shows channel maps integrated for the high-velocity $(|\Delta V|>2.0\ \kms$) compact emission, and Fig. \ref{fig:ce}b shows those integrated for the low-velocity ($|\Delta V|<1.0\ \kms$) extended emission where $\Delta V$ indicates the relative velocity to the systemic velocity. The maps of the high-velocity component show a very compact structure as a natural consequence of the integration of compact emission seen in the original channel maps shown in Fig. \ref{fig:ch}.
The dashed line in Fig. \ref{fig:ce}a passing the two peaks of blue- and redshifted components gives us a direction of the velocity gradient at the high velocities (P.A.$=72^{\circ}$), which is almost perpendicular to the outflow direction.
In contrast to the high-velocity component, the maps of the low-velocity component are extended to $\sim800$ AU scale at the 4$\sigma$ level. 
This figure also shows a clear velocity gradient in the direction of northeast-southwest as shown with a dashed line (P.A.$=49^{\circ}$) passing the two peaks of blue- and redshifted components in Fig. \ref{fig:ce}b. This direction of the velocity gradient of the low-velocity component is not perpendicular to the outflow axis and apparently different from that of the velocity gradient seen in the high-velocity component. These differences seen between the high-velocity and low-velocity components will be discussed in more detail later.

The ratio of the brightness temperature between the C$^{18}$O and $^{12}$CO emission tells us that C$^{18}$O is most likely optically thin although it is difficult for us to estimate the optical depth of the C$^{18}$O emission at $V_{\rm LSR}$ between $5\ \kms$ and $7\ \kms$, where $^{12}$CO is completely resolved-out. We note that when the ratio between the C$^{18}$O and $^{12}$CO emission is calculated, the spatial filtering has to be taken into account. In the case discussed here, however, the $^{12}$CO emission is significantly more extended than the C$^{18}$O emission, and as a result, the $^{12}$CO emission should be resolved out more than the C$^{18}$O emission. This suggests that the ratio of the C$^{18}$O emission to the $^{12}$CO emission should be overestimated, and as a result, the optical depth of the C$^{18}$O emission should be overestimated.
When C$^{18}$O is optically thin and under the condition of Local Thermodynamic Equilibrium (LTE), total gas mass is estimated from total flux of C$^{18}$O. The peak intensity and the total flux of the C$^{18}$O emission are $842\pm 82\ \mJB\,\kms$ and $7.20\pm 0.70\ {\rm Jy}\,\kms$, respectively. The gas mass calculated with the total flux is $M_{\rm gas}=4.4\times 10^{-3}\ \Ms$ assuming the excitation temperature $T=28$ K (Chandler et al. 1998) and the abundance of C$^{18}$O relative to H$_{2}$ of $X({\rm C^{18}O})=3.0\times 10^{-7}$ \citep{fr1982}. The gas mass derived from the C$^{18}$O flux is one order of magnitude smaller than that from dust continuum flux density, $4.2\times 10^{-2}\ \Ms$ assuming gas/dust=100 \citep[see a result for Class $\II$ disks by][]{wi.be2014}, though both mass estimations include a lot of uncertain factors such as optical depth, temperature, non-LTE effects, and gas/dust {\rm mass} ratio. Additionally the molecular abundance of C$^{18}$O might be decreased because of freeze-out of CO onto dust grains.

\section{Analysis}
\label{sc:ana}

\subsection{C$^{18}$O Velocity Gradient}
\label{sc:velgra}
As shown in the previous section, the C$^{18}$O emission arising mainly from the flattened envelope of the TMC-1A protostar shows a very clear velocity gradient. In this section, the nature of the velocity gradient is investigated in more detail.

\begin{figure}
\epsscale{1.0}
\plotone{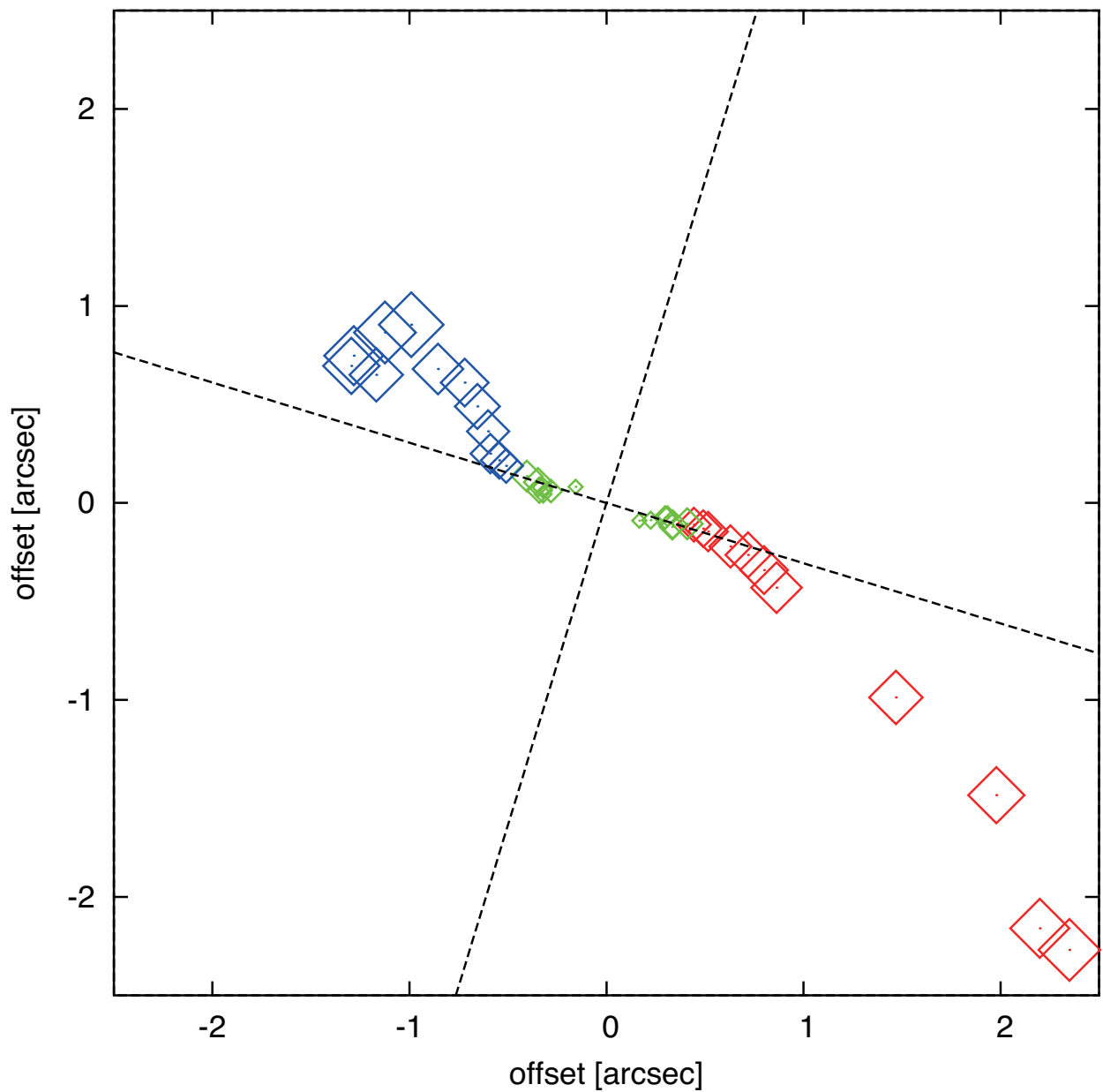}
\caption{2D mean positions of the C$^{18}$O $(J=2-1)$ emission in TMC-1A. Each point indicates a 2D mean position at each channel weighted with the intensity and $6\sigma $ cutoff ($\sigma$ is the rms noise level of C$^{18}$O emission). Error bars are not plotted but are smaller than the point size. Dashed lines show the major and the minor axis of the disk ($73^{\circ},\ -17^{\circ}$ respectively) and the intersection is the position of the central protostar (continuum emission peak). The point size becomes smaller as the value of $|V-V_{\rm sys}|$ becomes higher and green points correspond to a velocity range of $|\Delta V|>2.4\ \kms$. \label{fig:mp}}
\end{figure}

One remarkable characteristic of the velocity gradient is that its direction changes between high and low velocities as shown in Fig. \ref{fig:ce}. In order for us to investigate such a tendency more systematically (channel by channel), the mean position of the C$^{18}$O emission at each channel is measured, and plotted in Fig. \ref{fig:mp}. The mean position is measured as ${\bm x}_{0}(v)=\int {\bm x}I({\bm x},v)d^{2}x/\int I({\bm x},v)d^{2}x$, where the sum is calculated with pixels having an intensity more than $6\sigma$ ($\sigma $ is the rms noise level of C$^{18}$O). For channel at $|\Delta V|< 0.5\ \kms$, the mean positions are not measured because very complex structures appear at those channels, such as the ``X shape''. Points corresponding to higher velocities are plotted with smaller point sizes in Fig. \ref{fig:mp} and green points correspond to $|\Delta V|>2.4\ \kms$. This figure demonstrates the difference between high and low velocities seen in Fig. \ref{fig:ce} more systematically, i.e., the mean positions for higher velocities are mostly along the major axis of the disk (dashed line, E-W) and near the center. Those for lower velocities depart from the major axis and are located far from the center. This displacement changes the direction of the velocity gradient between higher and lower velocities. Fig. \ref{fig:mp} suggests that the motion in the inner (high-velocity) part is dominated by rotation of the disk because the velocity gradient is almost along the major axis of the disk. In contrast to the motion in the inner part, the motion in the outer (low-velocity) part is considered to include additional motions to rotation, which give a tilt in the velocity gradient with respect to the disk major axis.

In order for us to examine velocity gradients along the disk major axis in more detail, the Position-Velocity (PV) diagram cutting along the disk major axis shown in Fig. \ref{fig:pv}a is inspected here. This figure shows that the velocity gets higher as the position gets closer to the central star. As we discussed above, the PV diagram along the disk major axis is naturally considered to represent a velocity gradient due to rotation of the disk and/or the disk-like flattened envelope. There is a feature of the so-called ``spin up'' rotation where emission peaks get closer to the central position (horizontal dashed line) as the velocity increases, while emission peaks depart from the center at low velocities. Keplerian rotation $(V_{\rm rot}\propto R^{-1/2})$ is one form of spin up rotation, and rotation that conserves its specific angular momentum $(V_{\rm rot}\propto R^{-1})$ also shows a similar spin-up feature. However, it is generally not easy to distinguish such rotation patterns by comparing them with the PV diagram along the disk major axis. Further analysis to determine the powers of rotational velocity will be presented in the following section.

\subsection{Rotation Profile}
\label{sc:rotpro}
In order for us to characterize the nature of the rotation, detailed analysis of the PV diagram is presented in this section. The nature of the rotation can be characterized by its rotation profile obtained from the PV diagram by measuring a representative position at each velocity channel in the diagram.
The representative position at each velocity channel is defined as 1D mean position weighted by the intensity i.e. $x_{m}(v)=\int xI(x,v)dx/\int I(x,v)dx$ where the sum is done with pixels having an intensity more than $6\sigma$ ($\sigma $ is the rms noise level of C$^{18}$O). The measured mean positions are superposed on Fig. \ref{fig:pv}a with error bars. The error bar along the direction of the position is derived by the fitting while one along the direction of the velocity is the velocity resolution of the C$^{18}$O observations. The mean position is not measured at low velocities ($|\Delta V|<1.2\ \kms$) where the emission shows extended structures which is unsuitable for this analysis. We also note that a similar analysis was performed by \citet{ye2013}.

\begin{figure}
\epsscale{1.0}
\plotone{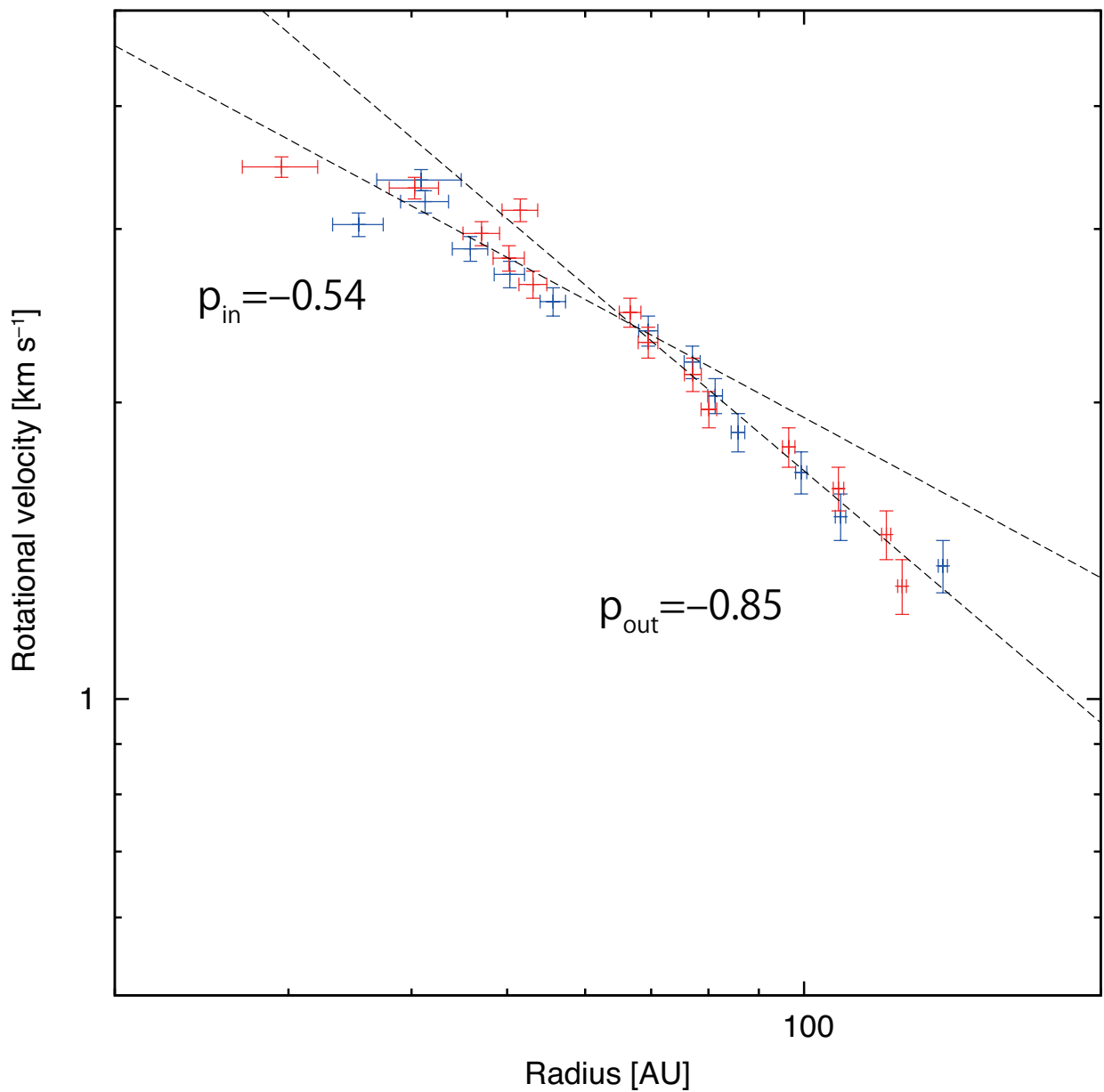}
\caption{Mean positions of the PV diagram along the major axis plotted on log$R-$log$V$ plane. The ordinate is not de-projected. Blue and red points show blueshifted and redshifted components, respectively. Dashed lines show the best-fit lines with a double power law. In addition to the inner and outer power, the ``break'' radius and velocity are included as free parameters. The best fit parameters are $R_{b}=67$ AU, $V_{b}=2.4\ \kms$, $p_{\rm in}=0.54\pm 0.14$, $p_{\rm out}=0.85\pm 0.04$. The error for $(R_{b},V_{b})$ is $\sim 1$\%.\label{fig:ll}}
\end{figure}

Fig. \ref{fig:ll} shows the mean positions of the PV diagram along the major axis on a $\log R-\log V$ diagram. The abscissa shows the mean positions measured as offset positions from the central protostar, which corresponds to the zero position on the PV diagram, assuming that the distance of TMC-1A is 140 pc. The ordinate is the relative-projected velocity assuming that the systemic velocity is 6.4$\ \kms$. 
Because there is no systematic separation between the blue- and red-shifted components, we consider that the central position and the systemic velocity are plausible in order to measure the relative position and velocity. Fig. \ref{fig:ll} shows clear negative correlation between radius and velocity, demonstrating that the rotation velocity can be expressed as a power-law function of the radius (spin-up rotation). Importantly, there seems to be a break point at $R\sim 70-80$ AU, where the power changes. We performed a $\chi^{2}$ fitting to Fig. \ref{fig:ll} with a double power model as follows:
\begin{eqnarray}
V_{\phi p}(R)=
\left\{
\begin{array}{c}
V_{b}(R/R_{b})^{-p_{\rm in}}\ \ \ \ (R\leq R_{b})\\
V_{b}(R/R_{b})^{-p_{\rm out}}\ \ \ \ (R>R_{b})\\  
\end{array}
\right.
\end{eqnarray}
The powers, $p_{\rm in}$ and $p_{\rm out}$, and the break point $(R_{b},V_{b})$ are free parameters. The best fit parameter set is $(R_{b},\ V_{b},\ p_{\rm in},\ p_{\rm out})=(67\ {\rm AU},\ 2.4\ \kms,\ 0.54\pm 0.14,\ 0.85\pm 0.04)$ where the errors for $R_{b}$ and $V_{b}$ are $\sim 1\%$, giving reduced $\chi^{2}=1.6$. Because this reduced $\chi^{2}$ is close to the unity, we consider that the best fit is reasonable. The error of each parameter is derived by dispersing the data points randomly within the error bars of $\log R$ and $\log V$. We adopted a Gaussian distribution for the error of $\log R$ and a rectangular distribution for the error of $\log V$ because the error of $\log V$ is based on the velocity resolution of the observations.
Another fitting to the mean positions is performed with a single power law. The best-fit parameter set for this single power law is $(R_{b},\ p)=(64\ {\rm AU},\ 0.69\pm 0.02)$ giving reduced $\chi^{2}=2.0$. In the single-power case, $V_{b}$ is fixed at 2.4 $\kms$ and $(R_{b},V_{b})$ does not give a break. Although more parameters usually make a fitting better or $\chi^{2}$ smaller, in our case, the best fit with a double power law provides a smaller ``reduced'' $\chi^{2}$ than that with a single power law even taking account the difference in the number of parameters between two fittings (four for the double power law and two for the single power law). This suggests that the rotation profile of TMC-1A can be fitted with the double power law better than the single power law. The best-fit result suggests that the inner part and the outer part follow different rotation laws. The power of the inner part is roughly equal to $1/2$ for the case of Keplerian rotation, suggesting that the inner/higher-velocity part is dominated by Keplerian rotation. By contrast, the outer/lower-velocity part, showing a slope steeper than Keplerian, cannot be supported by rotation against the gravity yielded by the central protostar, as we will discuss later. It is interesting to note that $p_{\rm out}$ is equal to $-0.85$, which is significantly different from $-1$ for the case where the specific angular momentum is conserved. This result could suggest that the specific angular momentum is proportional to $r^{0.15}$.

By using the break point $(R_{b},\ V_{b})=(67\ {\rm AU},\ 2.4\ \kms)$, the dynamical mass of the central protostar and the specific angular momentum at $R_{b}$, which are not inclination corrected, can be calculated at $M_{*p}=M_{*}\sin ^{2}i =0.43\ \Ms$ and $j_{p}=j\sin i=7.7\times 10^{-4}\ \kms\, {\rm pc}$, respectively. 
We should note, however, that the analysis using certain representative points on a PV diagram presented here may systematically underestimate $R_{b}$ when the spatial resolution is not high enough, as discussed in Appendix \ref{sc:app1}. Taking this point into consideration, the real break radius can be estimated to be $\sim 90$ AU from the apparent break radius of 67 AU. $R_{b}\sim 90$ AU leads $M_{*p}\sim 0.6\ \Ms$ and $j_{p}\sim 1\times 10^{-3}\ \kms\,{\rm pc}$.
\subsection{Rotating Disk Models}
\label{sc:rotmod}
In the previous section, the detailed analysis of the PV diagram cutting along the disk major axis was presented to characterize the nature of rotation, which suggests the existence of a Keplerian disk in the innermost part of the envelope. In this section, physical conditions of the Keplerian disk are investigated based on model fittings to the C$^{18}$O channel maps. To quantify physical parameters of the disk around TMC-1A, we performed a $\chi^{2}$ fitting to the C$^{18}$O $J=2-1$ channel maps based on a standard rotating disk model \citep[e.g.][]{du1994}. When a standard disk model is compared with these observed C$^{18}$O channel maps, it is important to note that the observed C$^{18}$O emission arises not only from the disk, but also from the envelope having rotation and additional motions as discussed in Sec.\ref{sc:velgra}. Because of this, it is required that only the velocity channels where rotation is dominant should be used for comparison with a disk model. As shown in Fig. \ref{fig:ll}, there is a velocity range showing the power-law index close to the Keplerian law, suggesting that the C$^{18}$O emission in this range with $|\Delta V|>2.4\ \kms$ is thought to arise from the possible Keplerian disk. We, therefore, intend to fit the channel maps within this velocity range, with rotating disk models. 

We use a disk model with a code described in \citet{oh2014}.
The model disk can be described  with 10 parameters summarized in Table \ref{tb:2}. 
The radial dependence of the disk temperature $T(R)$ and the disk surface density  $\Sigma (R)$ are described as
\begin{eqnarray}
T(R)&=&T_{100}\left( \frac{R}{100\ {\rm AU}}\right) ^{-q}\\
\Sigma (R)&=&\frac{(2-p)M_{200}}{2\pi(R_{\rm out}^{2-p}-R_{\rm in}^{2-p})}R^{-p}\ \ (R<R_{\rm out})
\end{eqnarray}
where $M_{200}$ is the mass within 200 AU ($R_{\rm out}$ will be fixed at 200 AU, see below for more detail). The scale height of the disk $H(R)$ is given under the hydrostatic equilibrium condition as follows:
\begin{eqnarray}
H(R)=\sqrt{\frac{2k_{B}T(R)R^{3}}{Gm_{0}M_{*p}/\sin ^{2}i}}
\end{eqnarray}
where $k_{B},\ G,\ m_{0}$, and $i$ are the Boltzmann constant, gravitational constant, mean molecular weight (=2.37 $m_{\rm H}$), and inclination angle of the disk, respectively. In such a disk, its density distribution $\rho (R,z)$ is given as $\Sigma(R)/(\sqrt{\pi}H(R))\exp(-z^{2}/H(R)^{2})$, while the rotation profile of the disk is provided by 
\begin{eqnarray}
V_{\phi}(R)=\sqrt{\frac{GM_{*p}/\sin ^{2}i}{R_{\rm cent}}}\left( \frac{R}{R_{\rm cent}}\right) ^{-v}
\end{eqnarray}
where $R_{\rm cent}$ is set to satisfy $V_{\phi}(R_{\rm cent})=\sqrt{GM_{*p}/\sin ^{2}i/R_{\rm cent}}$ and gives a coefficient of the power law of $V_{\phi}(R)$. 
It is assumed that the turbulent linewidth is zero; that is, the profile function is defined by a thermal linewidth. 

Among the 10 parameters, the first three parameters ($i$, $R_{\rm in}$, $R_{\rm out}$) are fixed, as shown in Table \ref{tb:2}. Note that $R_{\rm out}$=200 AU is assumed to cover a sufficiently large area for the fitting because our fitting to only the high-velocity inner part cannot constrain $R_{\rm out}$ well. In addition, $i=65^{\circ}$ is also suitable to explain motions in the envelope as discussed later in Sec.\ref{sc:inf}.  With the other seven free parameters $(M_{*p},M_{200},p,T_{100},q,R_{\rm cent},v)$, our model solves radiative transfers in 3D $+$ velocity space to produce model channel maps (data cube). When solving the line transfers, we assume Local Thermodynamic Equilibrium (LTE) and a typical abundance of C$^{18}$O relative to H$_{2}$, $X($C$^{18}$O$)=3.0\times 10^{-7}$. After solving the line transfers, model data cubes are convolved with the Gaussian beam, which has the same major axis, minor axis, and orientation as the synthesized beam of our observations. When the model is compared with the observations, the central position and the orientation (position angle, P.A.) of the disk are assumed to be the same as those of the continuum emission and the systemic velocity is also assumed to be $V_{\rm sys}=6.4\ \kms$.

Reduced $\chi^{2}$ is calculated to evaluate the validity of each model in the velocity range, $|\Delta V|> 2.4\ \kms$. We use only pixels in channel maps (data cube) where observed emission is stronger than the $3\sigma$ level because, if undetected pixels are included, where noise is distributed around $1\sigma$ (reduced $\chi^{2}\sim 1$), then the $\chi^{2}$ value would be underestimated. 
We use Markov Chain Monte Carlo (MCMC) method to find the minimum $\chi^{2}$ efficiently.

\begin{figure}
\epsscale{1.0}
\plotone{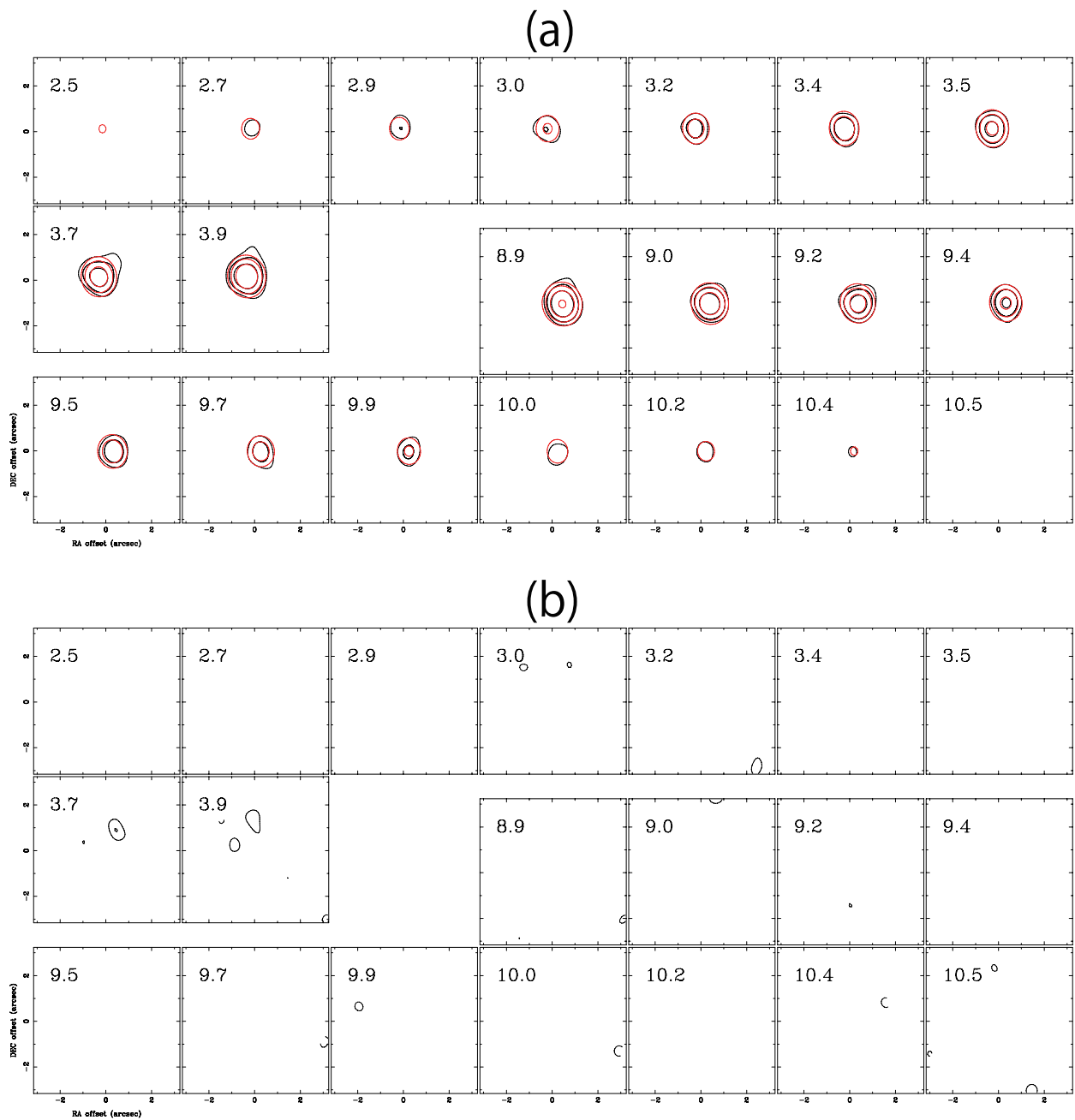}
\caption{Channel maps, used in a $\chi ^{2}$ fitting with the rotating disk models of (a) the best-fit disk model (red contours) superposed on that of the observations of C$^{18}$O $(J=2-1)$ emission and (b) the residual obtained by subtracting the best-fit disk model from the observations. The systemic velocity is $V_{\rm LSR}=6.4\ \kms$. Contour levels are $-3,3,6,12,24,...\times \sigma$ where 1$\sigma$ corresponds to $7.1\ \mJB$ for the panel (a) and $-3,-2,2,3,4...\times \sigma$ for the panel (b). The spatial scale is different from Fig. \ref{fig:ch}.\label{fig:modrot}}
\end{figure}

Fig. \ref{fig:modrot}a shows the comparison of the best-fit disk model with the observations and Fig. \ref{fig:modrot}b shows the residual obtained by subtracting the best-fit disk model from the observations. Reduced $\chi^{2}=0.72$ and Fig. \ref{fig:modrot}b indicate that our best-fit disk model reproduces, overall, the observations in the high-velocity region. 
The parameters of our best-fit disk model are summarized in Table \ref{tb:2}. First of all, $v=0.53$ suggests that the high-velocity component of C$^{18}$O emission from TMC-1A can be explained as a Keplerian disk better than other rotation laws, such as $V_{\phi}\propto R^{-1}$. Note that this power-law index is very consistent with $p_{\rm in}$ derived from the power-law fitting to the $\log R-\log V$ diagram in Sec.\ref{sc:rotpro}. The central stellar mass (not inclination corrected) $M_{*p}=M_{*}\sin ^{2}i=0.56\ \Ms$ is also consistent with that derived in Sec.\ref{sc:rotpro}. With the best-fit $M_{*p}$, $v$, and $R_{\rm cent}=166$ AU, the rotational velocity of our best-fit model is expressed as $1.5\ \kms (R/200\ {\rm AU})^{-0.53}$. The mass within $R_{\rm out}=200\ {\rm AU}$, $3.6\times 10^{-3}\ \Ms$ is similar to the total gas mass derived from the flux of C$^{18}$O, $M_{\rm gas}=4.4 \times 10^{-3}\ \Ms$. With the best-fit $M_{200}$ and $p=1.46$, the surface density profile of our best-fit disk model is expressed as $0.069\ {\rm g}\ {\rm cm}^{-2}(R/200\ {\rm AU})^{-1.46}$. The temperature seems roughly uniform ($q\simeq 0$) in our best-fit disk model. The temperature at the midplane can show such a radial profile when the inner region has a larger optical depth enough to cause a more effective cooling and a less effective heating by hot surface regions \citep{ch1997}. Another possibility is that the C$^{18}$O line traces a temperature at which C$^{18}$O molecules are evaporated off of dust grains.

The uncertainty of each parameter is defined by the range of the parameter where the reduced $\chi^{2}$ is below the minimum plus one ($=1.72$) when varying the parameter and fixing other parameters at those of the best-fit disk model. $M_{*}$ and $v$, which are related to kinematics, have smaller uncertainty as compared to others. Another parameter related to kinematics, $R_{\rm cent}$, has a large uncertainty because the rotational velocity is found to be close to the Keplerian rotation law. 

\begin{figure}
\epsscale{1.0}
\plotone{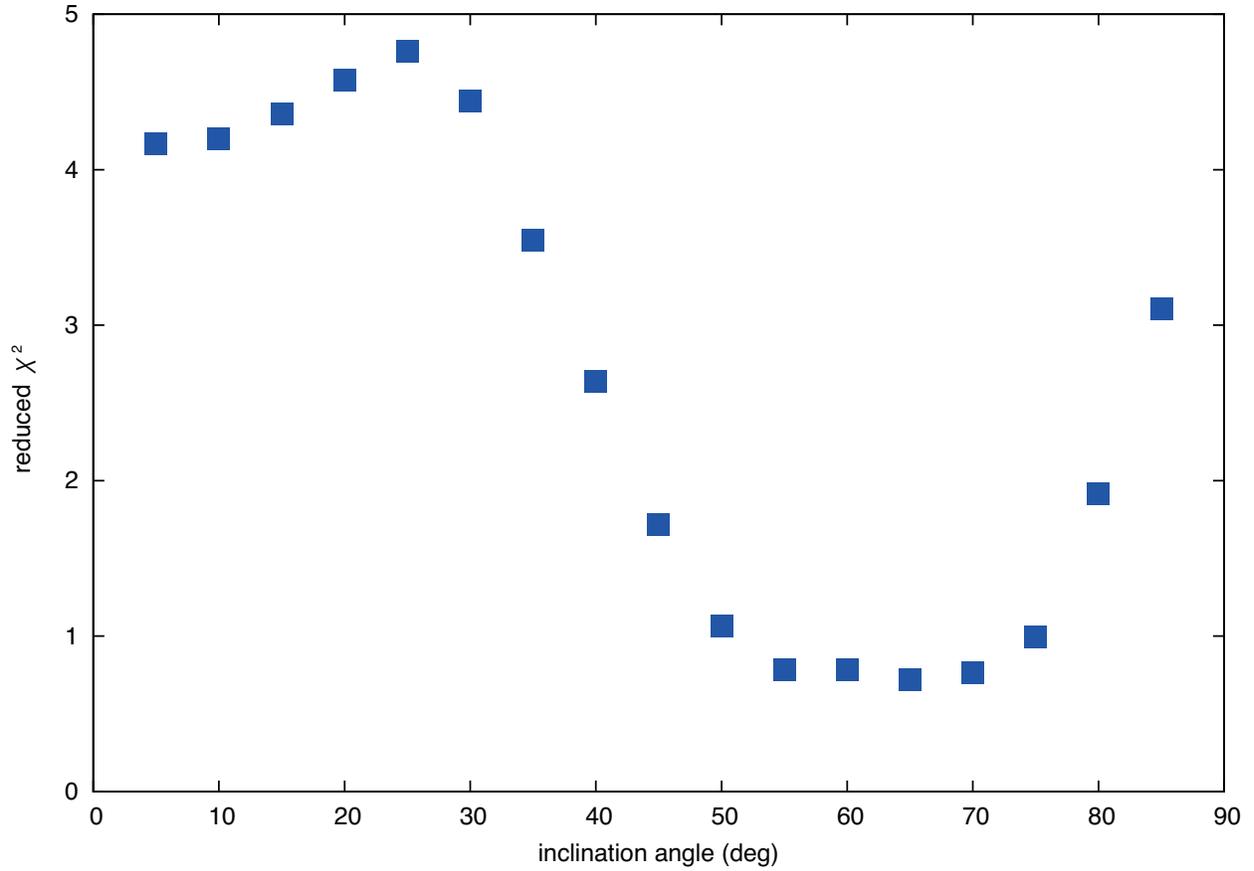}
\caption{Distribution of the reduced $\chi^{2}$ in terms of the inclination angle of the disk. All points have the same parameter set as the best-fit disk model except for the inclination angle. Larger inclinations ($45^{\circ}\leq i\leq75^{\circ}$) give small reduced $\chi ^{2}$ (that is, good) models to explain the observations.\label{fig:inc}}
\end{figure}

It is important to investigate the effect of the inclination angle, which is fixed at $i=65^{\circ}$ in our $\chi^{2}$ fitting mentioned earlier. Fig. \ref{fig:inc} shows a distribution of the reduced $\chi^{2}$ with different inclination angles of the disk. This plot is derived by changing the inclination angle $i$ from $5^{\circ}$ to $85^{\circ}$ by $5^{\circ}$ intervals, with the other parameters fixed at those of the best-fit disk model. Fig. \ref{fig:inc} shows that larger inclination angles between $45^{\circ}$ and $75^{\circ}$ give small reduced $\chi^{2}$, suggesting that it is difficult to give a constraint on the inclination angle in this model fitting. In Sec. \ref{sc:inf}, we examine the disk inclination angle further.

\section{Discussion}
\label{sc:dsc}
\subsection{Velocity gradient along the disk minor axis}
\label{sc:vgmi}
In the previous sections, detailed analysis of the motions for the inner part of the C$^{18}$O envelope where rotation is dominant are presented, showing that there is a Keplerian disk at the innermost envelope. As shown in Sec. \ref{sc:velgra}, the kinematics of the outer part of the envelope surrounding the Keplerian disk is described with not only rotation but also additional motions. In this section, the nature of the additional motions in the outer part of the envelope is discussed in detail.

There are two possibilities for the additional motions seen in the outer part. One is radial 
motion in the disk plane and the other one is motion perpendicular to the disk plane. Either case is expected to form velocity gradients along the disk minor axis. In the latter case, outflowing motion going perpendicularly to the disk plane is the most natural one. On the other hand, in the former case, infalling motion or expanding motion in the disk plane can be considered, but these two can be distinguished from each other because the directions of their velocity gradients are opposite to each other. The geometry of the outflow having its blueshifted lobe located on the northern side of the central star tells us that the northern part of the disk plane is on the far side from the observers, while the southern part is in the near side. Since the northern part of the C$^{18}$O emission is expected to have blueshifted emission as mentioned in Sec. \ref{sc:18}, the radial motion in the disk plane should be infall rather than expansion. Based on these considerations, the additional motion would be either an infall motion in the disk plane, an outflow motion going perpendicularly to the disk plane, or both.

\begin{figure}
\epsscale{1.0}
\plotone{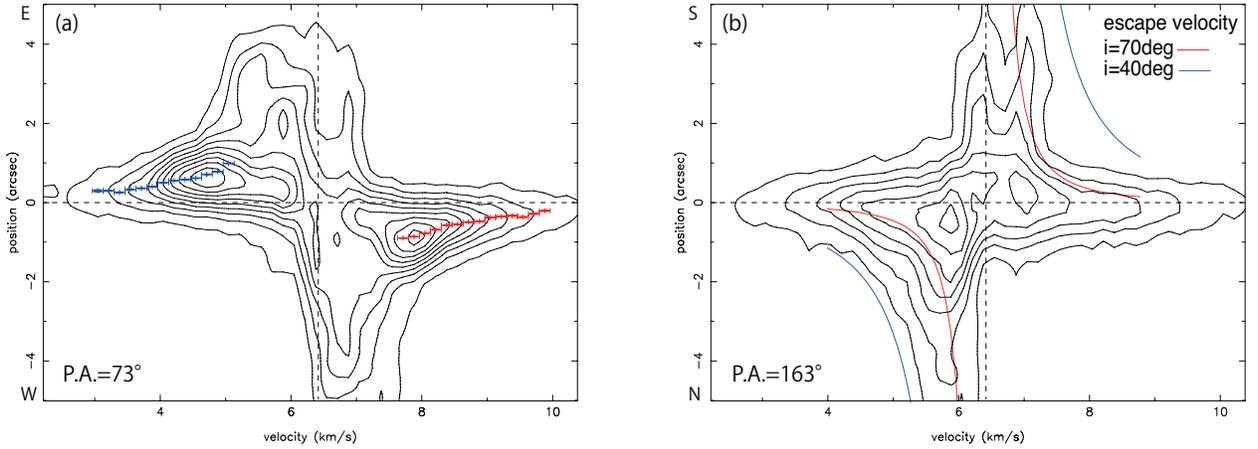}
\caption{Position-Velocity diagrams of the C$^{18}$O $(J=2-1)$ emission in TMC-1A along (a) the major and (b) the minor axes (the major axis corresponds to the white dashed line in Fig. \ref{fig:con}, PA $=73^{\circ}$). These PV diagrams have the same angular and velocity resolutions as those of the channel maps shown in Fig. \ref{fig:ch}. Contour levels are $6\sigma $ spacing from $3\sigma $ ($1\sigma =7.1\ \mJB$). Central vertical dashed lines show the systemic velocity ($V_{\rm sys}=6.4\ \kms$) and central horizontal dashed lines show the central position. Points with error bars in the panel (a) are mean positions derived along the position (vertical) direction at each velocity. Curves in the panel (b) show escape velocities along the outflow-axis direction for $i=70^{\circ}$ and $40^{\circ}$. \label{fig:pv}}
\end{figure}

The PV diagram along the disk minor axis shown in Fig. \ref{fig:pv}b shows a velocity gradient with blueshifted and redshifted components located on the northern side and the southern side, respectively at low velocities ($4.0\ \kms \lessim V\lessim 8.8\ \kms$). On the other hand, high-velocity ($V\lessim 4.0\ \kms,\ 8.8\ \kms \lessim V$) components show no significant velocity gradient along the disk minor axis, which is consistent with what was discussed with Fig. \ref{fig:mp}. 
The velocity gradient seen at low velocities can be explained as either infall motion in the disk plane, outflow motion going perpendicularly to the disk plane, or both, as discussed above. 
It should be noted that a molecular outflow needs velocities sufficiently high enough to escape from the gravitational potential due to a central star. Such escape velocities are estimated as a function of the distance from the central star and are potted in Fig. \ref{fig:pv}b.
When plotting the escape-velocity curves at the low velocities where there is a velocity gradient, the central mass (not inclination corrected) $M_{*p}=M_{*}\sin ^{2}i=0.56\ \Ms$ was assumed based on the estimation of the dynamical central stellar mass (see Sec. \ref{sc:rotmod}). The inclination angle was also assumed to be $40^{\circ}<i<70^{\circ}$ based on \citet{ch1996}. Compared with the curve with $i=70^{\circ}$, a part of the emission shows higher velocities than the escape velocities, suggesting an outflow motion. In fact, a part of the emission shown in channel maps (Fig. \ref{fig:ch}) exhibits extensions or X-shaped structures, which spatially correspond to the molecular outflow traced by $^{12}$CO. The significant part of the emission including the two peaks in Fig. \ref{fig:pv}b, however, show clearly slower velocities than the escape velocities. This suggests that the major part of the velocity gradient along the minor axis cannot be explained by outflow motions.
We therefore consider that the velocity gradient along the disk minor axis in C$^{18}$O emission is mainly due to infall motion in the disk plane. Note that additional signature for the infall motion can be also seen in the major-axis PV diagram at lower velocities ($5.5\ \kms \lessim V\lessim 7.0\ \kms$): although the most of the blueshifted emission is located on the eastern side, a part of the blueshifted emission at lower velocities is located on the western side, and similarly a part of the redshifted emission at lower velocities is located on the side opposite to the western side where the most of the redshifted emission is seen. These additional components in the diagram cannot be explained by rotation, and can be explained by infall \citep[e.g.][]{oh1997}. In addition, we also note that L1551 IRS5, a typical Class I protostar, also shows an infalling envelope with a velocity gradient along its disk minor aixs \citep{mo1998}. In the following subsections, details of infall motions seen in C$^{18}$O are discussed. 

\subsection{Nature of the infall motion}
\label{sc:inf}
It is suggested that the inner high-velocity part of the C$^{18}$O emission arises from a Keplerian disk and the parameters of the disk are derived by model fittings in Sec. \ref{sc:rotmod}. In addition to the rotation-dominant part, our results suggest that there is an infalling envelope surrounding the rotating disk. In order to investigate infall motions in the envelope, we make another model by adding an infalling envelope to the standard rotating disk model used in Sec. \ref{sc:rotmod}. Although the disk is basically described based in the best-fit disk model obtained in Sec. \ref{sc:rotmod}, it is necessary to define the transition radius between the disk and the infalling rotating envelope because the outer radius in the best-fit disk model is simply assumed to be 200 AU. A radius $R_{\rm kep}$ is set as a boundary between the infalling rotating envelope and the purely rotating disk. Within $R_{\rm kep}$, all the parameters are the same as the best-fit disk model obtained in Sec. \ref{sc:rotmod}, except for the inclination angle ($i$) which was not fully determined in Sec. \ref{sc:rotmod}. The velocity field of outer part $(R>R_{\rm kep})$ is described as below:
\begin{eqnarray}
V_{\phi }(R)&=&V_{\phi}(R_{\rm kep})\left( \frac{R}{R_{\rm kep}}\right)^{-v_{\rm out}}\\
V_{r}(r)&=&\alpha \sqrt{\frac{2GM_{*p}/\sin ^{2}i}{r}}
\end{eqnarray}
Note that we distinguish the axial radius $R=\sqrt{x^{2}+y^{2}}$ and the spherical radius $r=\sqrt{x^{2}+y^{2}+z^{2}}$ by using big and small letters, respectively. The infall-velocity vector points to the center and the infall velocity $V_{r}(r)$ depends on $z$ while the rotation velocity $V_{\phi}(R)$ does not depend on it. $V_{\phi}(R_{\rm kep})$ is the rotation velocity at $R_{\rm kep}$ and is assumed to connect continuously to that of the inner disk. The infall velocity is set as a product of a constant coefficient $\alpha$ and the free fall velocity. 
In terms of the geometry of the infalling envelope, we basically consider a flattened structure because of the shape of the C$^{18}$O integrated-intensity map (Fig. \ref{fig:mom}). This flattened geometry of the infalling envelope is modeled by extrapolating the density and temperature structures from the inner disk $(R<R_{\rm kep}$) to the outer radius of 900 AU, which is larger than the size of the C$^{18}$O integrated-intensity map. Hereunder, the models with a flattened envelope are called FE model.

It is important to note that in Sec. \ref{sc:rotmod}, the rotating disk models are fitted to the C$^{18}$O channel maps at higher velocities, while such fittings are not performed here. This is because models including the infalling envelope must be compared with the channel maps not only at higher velocities, but also at lower velocities with complicated asymmetric structures. Such structures cannot be fitted by our simple axisymmetric models. For this reason, in this section, we examine whether our models can reproduce the observed PV diagrams cutting along the major and minor axes (Fig. \ref{fig:pv}), by comparing the following two parameters between them: (1) the mean position in the major-axis PV diagram measured in Section \ref{sc:rotpro} (Fig. \ref{fig:pv}a and Fig. \ref{fig:ll}) and (2) the position of two peaks in the minor-axis PV diagram (Fig. \ref{fig:pv}b). The first depends on the rotation velocity while the second depends on the infall velocity within the disk plane. 

It is also noted that the surface density of the envelope at radii more than 200 AU is reduced by a factor of two to reproduce relatively diffuse emission seen in PV diagrams at positions larger than $\sim 2\arcsec$. Because extended structures more than $\sim 8\arcsec$ are resolved out in our ALMA observations by 50\%, we use the CASA task $simobserve$ to include the spatial filtering. The synthesized beam derived with $simobserve$ ($1\farcs 03\times 0\farcs 81,\ -7.2^{\circ}$) is not exactly the same as that of the observations even though the same antenna positions and hour angles are configured. However, the difference is small and there is no impact on our analysis.

Four parameters of the model mentioned above, $v_{\rm out}$, $R_{\rm kep}$, $\alpha $, and the inclination angle of the disk $i$ are changed to reproduce the observed PV diagrams. Fig. \ref{fig:modinf} shows the comparison of different models (FE model A-F) and the observations. The FE model A is the case with $(v_{\rm out},\ R_{\rm kep},\ \alpha,\ i)=(0.85,\ 100\ {\rm AU},\ 0.3,\ 65^{\circ})$. This FE model A can reproduce both PV diagrams, whereas the other five models cannot reproduce either of them as explained in detail below. Hereafter, we call the FE model A ``best-infall model''. We emphasize that the infall velocity is significantly smaller than the free fall velocity in the best-infall model. More details of the slow infall velocity is discussed later. 

FE model B and C, which have a shallower $v_{\rm out}$ and a smaller $R_{\rm kep}$, respectively, as compared with FE model A, show the clear difference in the $\log R-\log V$ diagram. The mean position derived from FE model B are located above the observations in $R\gtrsim 70$ AU. On the other hand, the mean positions of FE model C clearly show smaller radii than the observations in $R\gtrsim 50$ AU. 

FE model B and C exhibit similar PV diagrams along the minor axis to that of FE model A. However, FE model D and E, which have a larger infall velocity and a smaller inclination angle, respectively, as compared with FE model A, show clear difference in the minor-axis PV diagram. The two peaks of FE model D shows a clear offset from the observations along the ``velocity'' axis in the PV diagram. In addition, the $\log R-\log V$ diagram of FE model D is significantly different from the observations, in the sense that velocities at radii less than $\sim 80$ AU become larger than the observations, which can be also seen in the major-axis PV diagram where the two peaks in the model have larger velocities than the observations.
On the other hand, the two peaks in the minor-axis PV diagram of FE model E shows a clear offset along the ``position'' axis in the PV diagram.

The fact that FE model B, where the outer and inner power-law indices are the same as 0.53, cannot reproduce the observations indicates that the power-law index of the rotational velocity in the outer region is steeper than that in the inner region.
The discrepancy between the observations and the FE model C with $R_{\rm kep}=70$ AU, at which the observed rotation profile appears to break, indicates that $R_{\rm kep}$ should be larger than the apparent breaking radius, as also discussed in Appendix \ref{sc:app1} (see also Sec. \ref{sc:rotpro}). The discrepancy between the observations and FE model D, where the infall velocity is free fall velocity, indicates that the infall velocity in the envelope should be smaller than the free fall velocity.
The fact that the FE model E with $i=50^{\circ}$ cannot reproduce the observations indicates that the inclination angle of the disk is larger than $i=50^{\circ}$. 
Though the inclination angle is not well constrained with the disk model, as shown in Sec. \ref{sc:rotmod}, it is constrained better in FE models because the peak positions in the minor-axis PV diagrams are sensitive to the inclination angle.

In addition, the power law dependence of the infall velocity is investigated. Although, in FE model A, a conventional infall velocity law with the power-law index $-0.5$ is assumed, it would be possible that infall motions are decelerated and cease at the boundary radius because the inner region shows Keplerian rotation and no infall motion. Thus, we make FE model F with the same parameters as FE model A except for the infall velocity, which is linearly decelerated $V_{r}(r)=a (r-R_{\rm kep})\ (R>R_{\rm kep})$ where $a$ is a constant coefficient and set to be $a=1.6\times 10^{-3}\ \kms \,{\rm AU}^{-1}$. Overall velocity structures seen in both PV diagrams made from FE model F seem to be consistent with the observations, even though the peaks in the minor-axis PV diagram of this FE model F are located at a smaller velocity than the observations and the two peaks are almost merged. This would suggest that the infall velocity may decelerate, although such a deceleration can happen only at the innermost region of an envelope.
 
From the parameters of the best-infall model i.e. FE model A, $(v_{\rm out},\ R_{\rm kep},\ \alpha ,\ i)=(0.85,\ 100\ {\rm AU},\ 0.3,\ 65^{\circ})$, the infall velocity and mass infall rate $\dot{M}$ can be estimated. First, the inclination-corrected central stellar mass is calculated at $M_{*}=0.68\ \Ms$ from the inclination $i=65^{\circ}$ and the best-fit $M_{*p}=M_{*}\sin ^{2}i=0.56\ \Ms$ (best-fit disk model in Sec.\ref{sc:rotmod}). Thus, from $\alpha =0.3$, the infall velocity is estimated at $V_{r}=0.74-1.04\ \kms$ at $R=100-200$ AU. Using the surface density of the best-infall model, the mass infall rate is estimated as
\begin{eqnarray}
\dot{M}=2\pi RV_{r}\Sigma =(1.5-3.0)\times 10^{-6}\ \Ms \,{\rm yr}^{-1}\ (R=100-200\ {\rm AU}).
\end{eqnarray}
This value is comparable with a typical mass infall rate of protostars \citep[e.g.][]{oh1997} and the mass flow rate $\dot{M}_{\rm flow}=1.5\times 10^{-6}\ \Ms\,{\rm yr}^{-1}$ of TMC-1A estimated by an observation of the molecular outflow (Chandler et al. 1996). 
The inclination-corrected central stellar mass $M_{*}=0.68\ \Ms$ and the boundary radius $R_{\rm kep}=100$ AU also give us a specific angular momentum at the outer radius of the Keplerian disk, $j=1.2\times 10^{-3}\ \kms \,{\rm pc}$, which is consistent with the specific angular momentum at 580 AU, $j=2.5\times 10^{-3}\ \kms\ {\rm pc}$ within a factor of two \citep{oh1997}. From $R_{\rm kep}=100$ AU, the disk mass can be calculated to be $M_{\rm disk}=\int \Sigma 2\pi RdR=2.5\times 10^{-3}\ \Ms$, as well.

One might wonder whether a spherical geometry of the infalling envelope could explain the observations better, including the slower infalling velocities, even though the envelope has a flattened structure as shown in Sec. \ref{sc:18}. We therefore consider models with a spherical envelope having bipolar cavities in addition to FE models discussed above as shown in Appendix \ref{sc:app3}. It is found that models with a spherical envelope cannot explain the observations better than FE models. More importantly even if we adopt a model with a spherical envelope, the infall velocity should be significantly slower than the free fall velocity. In addition, it might be possible that power-law index $q$ for the temperature distribution in an envelope is not the same as that in a disk. A smaller value of $q$, i.e., $-0.5$ is examined in the model and found that it does not significantly change the results discussed above.

\subsection{Magnetic Pressure and Tension}
\label{sc:mag}
In previous sections, we suggest that the infall velocity in the envelope around TMC-1A is $\sim 0.3$ times as large as the free fall velocity, $V_{\rm ff}$, yielded by the stellar mass, which is estimated from the Keplerian rotation. The slow infall velocity suggests a possibility that infalling material are supported by some kind of force against the pull of gravity. One possible mechanism is the magnetic effect. Dense cores where protostars form tend to have oblate shapes perpendicular to their associated magnetic fields. \citep[e.g. NGC 1333 IRAS 4A,][]{gi2006}. Thus, it is assumed that envelopes around protostars should also be penetrated by magnetic fields, and the magnetic force (pressure and tension) possibly slows infall motions. In this section, we discuss whether the magnetic field expected to be associated with the envelope around TMC-1A can actually make the infall velocity as small as 0.3 $V_{\rm ff}$. Note that all the physical quantities in the following discussion are derived at a radius of 200 AU, which is rather arbitrary. The same estimations are also done at radii of 100 AU and 300 AU, providing results 
similar to that at 200 AU.

The magnetic flux density required to make the infall velocity as small as $0.3 V_{\rm ff}$ is estimated from the equation of motion including magnetic fields described below under a symmetric condition of $\partial /\partial \theta =0$ and $\partial/\partial z=0$ on the midplane: 
\begin{eqnarray}
\rho(R) V_{r}(R)\frac{dV_{r}}{dR}=-\frac{GM_{*}\rho(R) }{R^{2}}-\frac{1}{2\mu _{0}}\frac{dB^{2}}{dR}+\frac{B(R)^{2}}{\mu _{0}R_{\rm curv}(R)}
\label{eq:magr}
\end{eqnarray}
where $\rho (R)$, $V_{r}(R)$, and $R_{\rm curv}(R)$ indicate radial dependence of the mass density and the infall velocity in the midplane of the envelope, whereas $R_{\rm curv}(R)$ is the radial dependence of the curvature of magnetic field lines. $\mu_{0}$ indicates the magnetic permeability in a vacuum. The second and third terms of the right hand side indicate the magnetic pressure and the magnetic tension, respectively. A steady state is assumed here and the gas and radiation pressures are ignored because they are too weak to support material against gravity, i.e., the thermal energy is smaller than the gravitational potential by orders of magnitude at $\sim 200$ AU).

Although we do not know the configuration of magnetic fields around TMC-1A, it is reasonable for our order estimation to assume that $R_{\rm curv}$ is roughly equal to the scale height $H$, which is a typical spatial scale in the disk-like envelope. We also assume $B\propto \rho ^{2/3}$, which corresponds to the case where gases spherically symmetrically collapse with magnetic fields. Then, with $\rho (R)$ and $V_{r}(R)$ derived from the best-infall model discussed in the previous section,
the magnetic flux density required for $V_{r}=0.3V_{\rm ff}$ is estimated to be $B\sim 2$ mG.
The magnetic flux density on the order of mG is somewhat larger than the typical value from Zeeman observations ($\lessim 100\ \mu $G). However, the Zeeman observations using a high-density tracer such as CN give us higher values, $\lessim 1$ mG \citep{cr2010}. In addition to single-dish observations, observations using SMA by \citet{gi2006} toward NGC 1333 IRAS 4A at $\sim 1\arcsec$ angular resolution estimated the magnetic field strength in the plane of the sky to be $B_{\rm POS}\sim 5$ mG by applying the Chandrasekhar-Fermi equation. These observations \citep[see also][]{fa2008,st2013} suggest that the magnetic field strength around TMC-1A can be possibly on the order of mG.

\subsection{A Possible scenario of the Keplerian Disk Formation}
Over the last decade the number of protostellar sources associated with Keplerian disks have been increased with (sub)millimeter interferometric observations (see Table 3), and it is getting more feasible for us to discuss evolutionary processes of Keplerian disks around protostars based on these observations \citep[e.g.][]{ch2014}. We should, however, note that disk radii measured from such observations are subject to serious ambiguity, because the measured radii are likely affected by the effects of the sensitivity limits and the missing fluxes of the interferometric observations. In particular, some of these observations did not identify the transition from infalling motions to Kepler motions (see Column 8 in Table 3), so that the estimated disk radii measured from such observations should be lower limits to the actual disk radii (see more details below). Indeed, \citet{ch2014} attempted to discuss evolutionary processes of Keplerian disks around protostellar sources with those samples, but they could not identify any clear correlations and evolutionary sequences of the disks.

On the other hand, our series of ALMA observations of protostellar sources, L1527 IRS by \citet{oh2014}, TMC-1A in the present work, and L1489 IRS by \citet{ye2014}, all identified the transitions from infalling envelopes to disks, and measured the power-law indices of the rotational profiles to verify that the disks are indeed in Keplerian rotation. In such observations, the disk radii are estimated better. 
Furthermore, recent high-resolution interferometric observations of L1551 NE \citep{ta2014}, L1551 IRS 5 \citep{ch2014}, and VLA1623A \citep{mu2013} have also revealed transitions from infalling motions to  Keplerian rotations. These observational results imply that transitions from infall motions to Keplerian rotations are ubiquitous in the protostellar phase. Statistical studies of such systems should provide important observational constraints on the ongoing growth process of Keplerian disks.
Here, with our ALMA results, along with the other observational results showing transitions from infalling envelopes to Keplerian disks, we re-examine evolutionary processes of Keplerian disks around protostellar sources embedded in infalling envelopes.

In the following discussion, the bolometric temperature ($T_{bol}$) and the ratio of the bolometric to submillimeter luminosity $(=L_{bol}/L_{\rm submm}$) are adopted as parameters tracing the evolution of protostars. \citep{gr2013} Fig. \ref{fig:rkms} shows correlations between $T_{\rm bol}$ vs the central protostellar masses ($=M_{*}$), $L_{bol}/L_{\rm submm}$ vs $M_{*}$, and $M_{*}$ vs the radii of the Keplerian disks ($=R_{\rm kep}$) for the protostellar sources associated with the Keplerian disks listed in Table 3. Note that the protostellar sources ``with transitions'' showing transitions from infall to Keplerian and ``without transitions'' are indicated with different marks. The sources with transitions exhibit correlations in all the $T_{bol}$ vs $M_{*}$, $L_{bol}/L_{\rm submm}$ vs $M_{*}$ and $M_{*}$ vs $R_{\rm kep}$ plots, with the correlation coefficients of 0.97, 0.95, and 0.91, respectively. The correlation coefficient for $M_{*}$ vs $R_{\rm kep}$ will be 0.98 if assuming that VLA1623A has $R_{\rm kep}=50$ AU, at which its rotational velocity profile shows a turn over point \citep{mu2013}. These results may suggest that the central protostellar masses increase as the protostellar evolution proceeds, and that the disk radii also grow as the central protostellar masses grow, although statistical significance are not high because the number of sample is small. On the other hand, inclusion of all the sample points almost diminishes the correlation in $M_{*}$ vs $R_{\rm kep}$ with the correlation coefficient of 0.29. This poor correlation may be because $R_{\rm kep}$ is not measured accurately for the cases without transitions. The correlation between $L_{bol}/L_{\rm submm}$ and $M_{*}$ is not diminished (the correlation coefficient is 0.77) because measuring $M_{*}$ does not depend on whether there is a transition though the correlation of $T_{bol}$ vs $M_{*}$ is poor when all the samples are included (the correlation coefficient is 0.43). These results demonstrate that proper observational estimates of the central protostellar masses and the disk radii are essential to discuss evolutions of the protostars and disks.

A simple interpretation of the positive correlation between $M_{*}$ and $R_{\rm kep}$ is given by a conventional analytical picture of disk formation with ``inside-out collapse'' of envelopes, where the specific angular momenta increase linearly at outer radii (i.e., rigid-body rotation) \citep{te1984,sh1987,ye2013}. At the onset of the collapse, only the central part of the envelope with less angular momentum falls toward the center, and thus the mass of the protostellar source and the radius of the Keplerian disk are smaller. As the collapse proceeds outward, more material with larger angular momenta will be carried into the central region, raising the protostellar mass and radius of the Keplerian disk. In this picture, the dependence of $R_{\rm kep}$ on $t$ (time) and $M_{*}$ can be expressed as shown in \citet{ye2013}:
\begin{eqnarray}
R_{\rm kep}=\frac{\omega ^{2}r_{\rm inf}^{4}}{GM_{*}(r_{\rm inf})}=\frac{\omega ^{2}G^{3}\dot{M}^{3}}{16c_{s}^{8}}t^{3}=\frac{\omega ^{2}G^{3}}{16c_{s}^{8}}M_{*}^{3}
\label{eq:rkms}
\end{eqnarray}
where $\omega$ and $c_{s}$ denote the angular velocity and the sound speed of the envelope, respectively, $r_{\rm inf}$ is the outermost radius of the infalling region, and $\dot{M}$ is the mass infalling rate. A typical value of $\omega$ has been estimated as $\omega \sim 1\ \kms \ {\rm pc}^{-1} = 3.2 \times 10^{-14}\ {\rm s}^{-1}$ \citep{go1993}. In Fig. \ref{fig:rkms}b lines of Eq.(\ref{eq:rkms}) with twice and half of the typical value of $\omega$ are drawn, with $c_{s}=0.19\ \kms$ (i.e., $T_{k} = 10$ K). Almost all the sources with transitions are within the area expected from Eq. \ref{eq:rkms}, suggesting that the identified positive correlation between $M_{*}$ and $R_{\rm kep}$ can be reproduced with the picture of the conventional analytical disk-formation model with inside-out collapse.

More realistically, effects of magnetic fields on disk formation and growth, which are ignored in the model described above, must be taken into account \citep{ma2014,li2014}. Our ALMA observations of TMC-1A and L1527 IRS have found that the infalling velocities are a factor of a few slower than the corresponding free fall velocities derived from the central protostellar masses, while in L1489 IRS the infalling velocity is consistent with the free fall velocity. As discussed in the previous section, the slow infall velocities might be due to magnetic fields, while in the case of L1489 IRS, the magnetic field may not be effective any more because the surrounding envelope that anchors the magnetic field is almost dissipated in the late evolutionary stage \citep{ye2014}. The latest theoretical simulations by \citet{ma.ho2013}, which include magnetic fields, have successfully demonstrated the growth of protostars and disks, as well as the deceleration of infall and the dissipation of envelopes in the late evolutionary stage. Their models also predict that, in the last phase of evolutionary stage, the growth of the central Keplerian disk stops while the protostellar mass itself is still growing \citep{vi2009,ma.ho2013,ma2014}. Currently our observational samples do not show any clear evidence for such a saturation of the disk radius. Future large-scale ALMA surveys of a number of protostellar sources may reveal such a saturation, and a comprehensive observational picture of disk formation and growth around protostellar sources.

\section{Conclusions}
\label{sc:cnc}
Using ALMA in Cycle 0, we observed a Class I protostar in Taurus star-forming region, TMC-1A in the 1.3-mm dust continuum, the $^{12}$CO ($J=2-1$), and C$^{18}$O ($J=2-1$) lines. The main results are summarized below.

1. The 1.3-mm dust continuum shows a strong compact emission with a weak extension to the west. The gas mass of the dust emission is estimated from its flux density to be $M_{\rm dust}=4.2\times 10^{-2}\ \Ms$ assuming a opacity $\kappa _{1.3{\rm mm}}=1.2\ {\rm cm}^{2}\,{\rm g}^{-1}$, $T_{c}=28$ K and the gas/dust mass ratio of 100, which is ten times larger than the gas mass estimated from the C$^{18}$O flux. 

2. The $^{12}$CO line traces the molecular outflow with the axis perpendicular to the elongation of the 1.3-mm continuum emission. Velocity gradients are seen both along and across the outflow axis. The velocity is accelerated along the outflow axis (Hubble law). Across the outflow axis, the higher velocity component is located closer to the axis, which can be explained by an outflow driven by a parabolic wide-angle wind.

3. The C$^{18}$O emission shows the velocity gradient along the major axis of the disk traced in the 1.3-mm continuum emission. This velocity gradient is due to rotation. The power-law indices of the radial profile of the rotation velocity are estimated from fittings to the major-axis PV diagram to be $p_{\rm in}=0.54$ for the inner/high-velocity component and $p_{\rm out}=0.85$ for the outer/low-velocity component. This indicates the existence of the inner Keplerian disk surrounded by the outer infalling envelope.

4. In order to investigate the nature of the Keplerian disk, $\chi ^{2}$ fittings are performed to the C$^{18}$O channel maps at high velocities with a rotating disk models. In the model fitting, the power-law index of the rotational velocity ($v$) is included as a free parameter. The reduced $\chi^{2}$ is minimized at $v=0.53$, which is very consistent with $p_{\rm in}$, and confirms that the rotational velocity at high velocities can be explained by Keplerian rotation. The dynamical mass (not inclination corrected) derived from our best-fit disk model is $M_{*}\sin ^{2}i=0.56\ \Ms$.

5. In addition to the velocity gradient along the major axis, there is another velocity gradient along the minor axis, which can be interpreted as infalling motions. In order to investigate the nature of the infall motion, 
the observed PV diagrams along both major and minor axes are compared with models consisting of an infalling envelope and a Keplerian disk. The Keplerian-disk size and the inclination angle are estimated to be 100 AU and $i=65^{\circ}$ respectively, and the infall velocity is found to be 0.3 times as large as the free fall velocity yielded by the dynamical mass of the central protostar. Possible reasons for the infall motions to be slower than the free fall velocity are magnetic pressure and magnetic tension. 
The magnetic field strength yielding the slow infall velocity is estimated to be $\sim 2 $ mG. 

6. Based on observations, including our series of ALMA observations, which show transitions from infalling envelopes to Keplerian disks, evolutionary processes of Keplerian disks around protostellar sources are examined. Those samples exhibit correlations in $T_{\rm bol}$ vs $M_{*}$, $L_{bol}/L_{\rm submm}$ vs $M_{*}$, and $M_{*}$ vs $R_{\rm kep}$. This may suggest that the central protostellar masses and disk radii increase as protostellar evolution proceeds. 

\acknowledgments
This paper makes use of the following ALMA data: ADS/JAO.ALMA\#2011.0.00210.S (P.I. N. Ohashi). ALMA is a partnership of ESO (representing its member states), NSF (USA) and NINS (Japan), together with NRC (Canada) and NSC and ASIAA (Taiwan), in cooperation with the Republic of Chile. The Joint ALMA Observatory is operated by ESO, AUI/NRAO and NAOJ. We thank all the ALMA staff making our observations successful. We also thank the anonymous referee, who gave us invaluable comments to improve the paper.

$Facilities:$ ALMA

\appendix
\section{Analytic Consideration on our Method to Derive Rotational Profiles from Observed PV Diagrams}
\label{sc:app1}
In section \ref{sc:rotpro} we adopt two independent methods to derive the rotational profile of the disk around TMC-1A;
one is a method using the observed PV diagram along the major axis introduced by \citet{ye2013},
and the other a full three-dimensional $\chi^{2}$ fitting of a rotating disk model
to the observed C$^{18}$O channel maps in the high-velocity ($>2.4$ $\kms$) regions.
The former method yields the rotational profile of $2.4\ (\kms) \times (R/67\ {\rm AU})^{-0.54\pm 0.03}$,
and the latter $2.4\ (\kms) \times (R/87\ {\rm AU})^{-0.53\pm 0.1}$.
Thus, while the derived rotational power-law indices
are consistent with each other and suggest Keplerian rotation,
the rotational radius at a certain velocity derived from the former method is $\sim 23\%$  lower
than that from the latter method. This implies that the former method underestimates the central stellar mass and the
rotational angular momentum by $\sim 23\%$.
In this appendix, these uncertainties inherent in the method using the PV diagrams
are analyzed.
For simplicity, we consider a geometrically-thin, uniform disk with a pure rotational motion,
and discuss the effects of the finite velocity and spatial resolutions.
The effect of the internal gas motion in the disk can be absorbed
into the effect of the finite velocity resolution. We do not consider any radial motion in the disk here,
since the effects of the radial motions on the estimates of the rotational profiles have already been
discussed by \citet{ye2013}.

The on-plane coordinates of the model disk are described with
the 2D polar coordinates system $(R,\phi)$, and the projected coordinates on the plane of the sky
2D cartesian $(x,y)$, where the $x$-axis is defined as the major axis of the ellipse of the disk
projected on the sky. The disk inclination angle from the plane of the sky ($\equiv i$) is defined
such that $i=0$ means face-on, and thus  $x= R\cos \phi$ and $y= R\sin \phi \cos i$.
The radial profile of the rotation velocity of the disk is expressed as $V_{\phi}(R)=V_{1}(R/R_{1})^{-v}$, and the iso-velocity curve at a given line-of-sight velocity ($\equiv V_{\rm LOS}$) projected onto the plane of the sky is described as $V_{\rm LOS}=V_{\phi}(R)\cos \phi \sin i$.
This curve passes through the origin $(	x,y) = (0,0)$ and $(R_{o},0)$ where $R_{o}=R_{1}(V_{\rm LOS}/V_{1}/\sin i)^{-1/v}$. $R_{o}$ gives the outermost point located in the major axis at a given $V_{\rm LOS}$ and curves at different $V_{\rm LOS}$ do not intersect with each other.
The curves exhibit a symmetrical shape with respect to the major axis, and around the systemic
velocity the curves show the well-known ``butterfly'' shapes.
The curves at different $V_{\rm LOS}$ correspond to the velocity channel maps
in the case of the infinite spatial and velocity resolutions,
and the PV diagram along the major axis forms the rotation curve
of $V_{\rm LOS}=V_{\phi}(R)\sin i$ exactly.
In cases of finite spatial and velocity resolutions, the iso-velocity curve on the plane of the sky
must have widths depending on the velocity resolution ($\equiv \Delta V$), where the maximum width
is seen along the major axis.
Such iso-velocity regions are convolved with the spatial beam, defined as the circular Gaussian
with the FWHM beam size of $\theta$ ($\equiv \exp [-4\ln 2(x^{2}+y^{2})/\theta ^{2}])$.
The spatial convolution skews the mean position along the major axis ($\equiv x_{m}$) at a given $V_{\rm LOS}$.

In realistic cases of central disks, our method adopts only high-velocity (typically $>$ a few times km s$^{-1}$),
fast-rotating parts of the P-V diagrams, and the amount of the positional shifts within a typical velocity
resolution ($\sim 0.1\ \kms$) is much smaller than a typical spatial resolution.
The iso-velocity curve ($i.e.$, $V_{\rm LOS}=V_{\phi}(R)\cos \phi \sin i$) can be re-written in terms of $y$ as
\begin{eqnarray}
\frac{y}{R_{o}\cos i}=\pm \sqrt{\left(\frac{x}{R_{o}}\right)^{2/(v+1)}-\left( \frac{x}{R_{o}}\right) ^{2}}.
\end{eqnarray}
Thus, the width of the iso-velocity region along the $y$ direction at a given $x(<R_{o})$ and $\Delta V$ can be written as
\begin{eqnarray}
\Delta y\simeq \frac{\partial y}{\partial V_{\rm LOS}}\Delta V=\frac{\partial y}{\partial R_{o}}\frac{dR_{o}}{dV_{\rm LOS}}\Delta V
=\frac{R_{o}^{v+1}\Delta V\cos i}{(v+1)V_{1}R_{1}^{v}\sin i}\frac{\left(\frac{x}{R_{o}}\right) ^{2/(v+1)}}{\sqrt{\left(\frac{x}{R_{o}}\right)^{2/(v+1)}-\left( \frac{x}{R_{o}}\right) ^{2}}}.
\end{eqnarray}
where $(\Delta V/V_{\rm LOS})^{2}$ or higher order components are ignored.
On the major axis ($y=0$), the intensity distribution ($\equiv I$) spatially convolved along
the $y$ axis is then described as;
\begin{eqnarray}
\label{eq:a3}
I_{\theta _{n}}(z),\propto \Delta y\exp \left[ -4\ln 2\left( \frac{y}{\theta }\right) ^{2}\right] \propto \frac{z^{\frac{1}{v+1}}}{\sqrt{1-z^{v+1}}}\exp \left[ -4\ln 2 \left(\frac{z^{\frac{1}{v+1}}\sqrt{1-z^{v+1}}}{\theta _{n}}\right) ^{2}\right ]
\end{eqnarray}
where $z=x/R_{o}$ is the normalized spatial coordinate along the major axis, and $\theta _{n}=\theta /R_{o}/\cos i$
is the normalized beam width along the minor axis.
In Eq.(\ref{eq:a3}), $\Delta y^{2}$ or higher order components are ignored because  $\Delta y$ is an infinitesimal proportional to $\Delta V$. The spatial convolution along $x$-axis (the major-axis direction) does not change the mean position mathematically. Then, the skewness of the mean position along the major axis due to the finite resolution $\Delta V$ and $\theta$ can be expressed as;
\begin{eqnarray}
z_{m}=\frac{x_{m}}{R_{o}}=\frac{\int_{0}^{1}I_{\theta _{n}}(z)zdz}{\int_{0}^{1}I_{\theta _{n}}(z)dz}
\end{eqnarray}
where $z_{m}$ is the normalized mean position.
$z_{m}$ can be derived as a function of $\theta _{n}$, which depends on $R_{o}$ (or $V_{\rm LOS}$) because $\theta $ is a constant.
Fig. \ref{fig:z0} shows the dependence of $z_{m}$ on $1/\theta _{n}$ in the case of $v=1/2$ (Keplerian rotation).
This figure shows that large $1/\theta _{n}$ gives $z_{m}\simeq 1$, and thus
the relevant radius can be derived accurately if the spatial resolution is high enough as compared with the emission size.
On the other hand, small $1/\theta _{n}$ gives $z_{m}\simeq 0.760<1$,
which means that the correct radius cannot be derived from the mean positions.
However, if the beam size is very large
($1/\theta _{n}<1.0$), the dependency of $z_{m}$ on $1/\theta _{n}$ (i.e. $R_{o}$) is small, indicating
$x_{m}=const.\times R_{o}$. This means the power-law index can be derived correctly by using the mean positions,
though the radii themselves are underestimated. 
These results also indicate that the estimates of the central stellar masses
or angular momenta are affected at most $<$23$\%$. 

In our case of TMC-1A, the size of the synthesized beam (FWHM$\sim 1\arcsec$) corresponds to $1/\theta _{n}\simeq R_{o}/1\farcs4=R_{o}/200\ {\rm AU}$ with $i=65^{\circ}$ and $d=140$ pc.
Fig. \ref{fig:z0} shows that the power-law index of the rotational velocity is expected to be estimated correctly
at $R<200$ AU for TMC-1A. In an intermediate situation, the degree of the underestimation
of the radii depends on the relevant radius, thus the power-law indices must also be distorted.
These considerations imply that effects of the spatial and velocity resolutions, as well as those of
the radial velocities, should be taken into account to derive the rotational profiles from the observed PV diagrams.

\section{Comparison between the flattened and spherical envelope models}
\label{sc:app3}
In Sec. \ref{sc:inf}, we have constructed models consisting of a Keplerian disk and a flattened envelope (FE model) to interpret the observed gas motions, and have found that the envelope is infalling toward the disk with the infalling velocity slower than the free fall velocity yielded by the mass of the central star. The interpretation of gas motions are, however, dependent on the assumed geometry of the envelope \citep[e.g.][]{to2012b}, and it is thus possible that the identified infalling motion slower than the free fall could be due to the effect of the assumed envelope geometry. To qualitatively assess the effects of the different geometries of the envelopes, we have also constructed models consisting of a spherical envelope and a Keplerian disk. Hereafter we call them SE models.

The density structure of the spherical envelope in the SE models we adopted is a solution of the rotating and infalling envelope modeled by \citet{te1984}, with the temperature profile of $T(r) = 38.0\ {\rm K}\ (r/100\ {\rm AU})^{0.02}$, the mass infalling rate of $\dot{M}=10^{-5}\ \Ms /{\rm yr}\ (M_{*}/0.5\ \Ms )^{0.5}$, the central stellar mass $M_{*} = 0.68 \Ms$, and the outermost radius $R=900 $ AU. To mimic the cavity created by the associated outflow, artificial bipolar cavities in a conical shape with the opening angle of 20$^{\circ}$ are added in the spherical envelope, and within $R < R_{\rm kep} = 100$ AU the same disk as that of FE models is located. 
To simplify the comparison of the observed gas kinematics between FE and SE models, the rotational velocity in the spherical envelope is assumed to be only the function of the distance from the rotational axis, with the same radial profile as that of the flattened envelope models, and the infalling motion is assumed to be isotropic. As the case of FE models, we adopt two types of infalls in SE models; a reduced infall $V_{r}=0.3V_{\rm ff}$, and free fall ($V_{r}=V_{\rm ff}$), where the free fall velocity is derived from the central stellar mass of $0.68\ \Ms$. The former and latter models are named as SE model A and D, respectively. The corresponding FE models are FE model A (our best fit model) and D.

Figure \ref{fig:saigo}a and b compare the PV diagrams along the major and minor axes of FE model A (red contours) and SE model A (grey scale), and Fig. \ref{fig:saigo}c and d compare FE model D (red contours) and SE model D (grey scale), respectively. These figures show some slight differences between SE and FE models as described below. As shown in Fig. \ref{fig:saigo}d, the minor-axis PV diagram of SE model D shows two peaks at $V\sim 5\ \kms$ and $\sim 8\ \kms$, while two velocity peaks in FE model D are $\sim 1\ \kms$ faster than them. In addition, FE model A/D shows only blueshifted emission at positions $\lessim -1 \arcsec$ in the minor-axis PV diagram and only redshifted emission at $\gtrsim 1\arcsec$, whereas SE model A/D shows blue- and redshifted emission on both positive and negative sides. In spite of these slight differences, FE model A and SE model A, as well as FE model D and SE model D, show overall similar PV diagrams along both major and minor axes and we, therefore, conclude that SE models cannot explain the observations better than FE models.
It is important to note that when SE model A and D are compared with the observations, it is found that SE model A can explain the observations better than SE model D, suggesting that the infall velocity should be significantly slower than the free fall velocity even in the case with a spherical envelope.

\clearpage

\bibliographystyle{apj}
\bibliography{reference_aso}

\clearpage

\begin{figure}
\epsscale{1.0}
\plotone{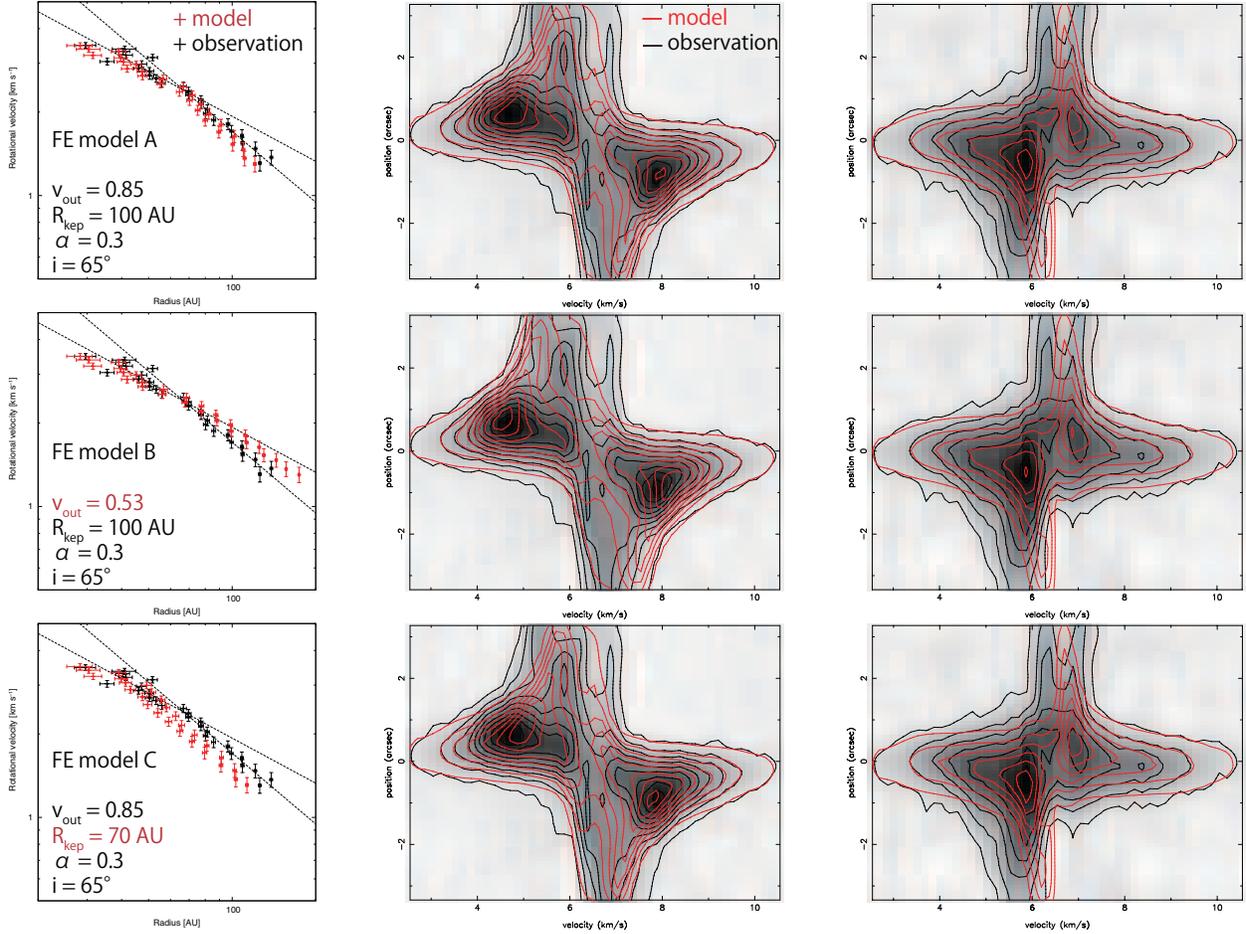}
\caption{Comparisons of the observations and models with infall motions. The left column shows the mean position of the PV diagram along the major axis in $\log R-\log V$ diagrams. Red points indicate models while black points indicate the observation, which is the same plot as Fig. \ref{fig:ll} except for the color. The middle and right columns show PV diagrams along the major and minor axis respectively. The observations are in black contours and grayscales while models are in red contours. Contour levels of the PV diagrams are the same as Fig. \ref{fig:pv}. The same row includes the same model and the parameters of each model are indicated in the left panel of each model. FE model A explains the observations most reasonably. \label{fig:modinf}}
\end{figure}

\addtocounter{figure}{-1}
\begin{figure}
\epsscale{1.0}
\plotone{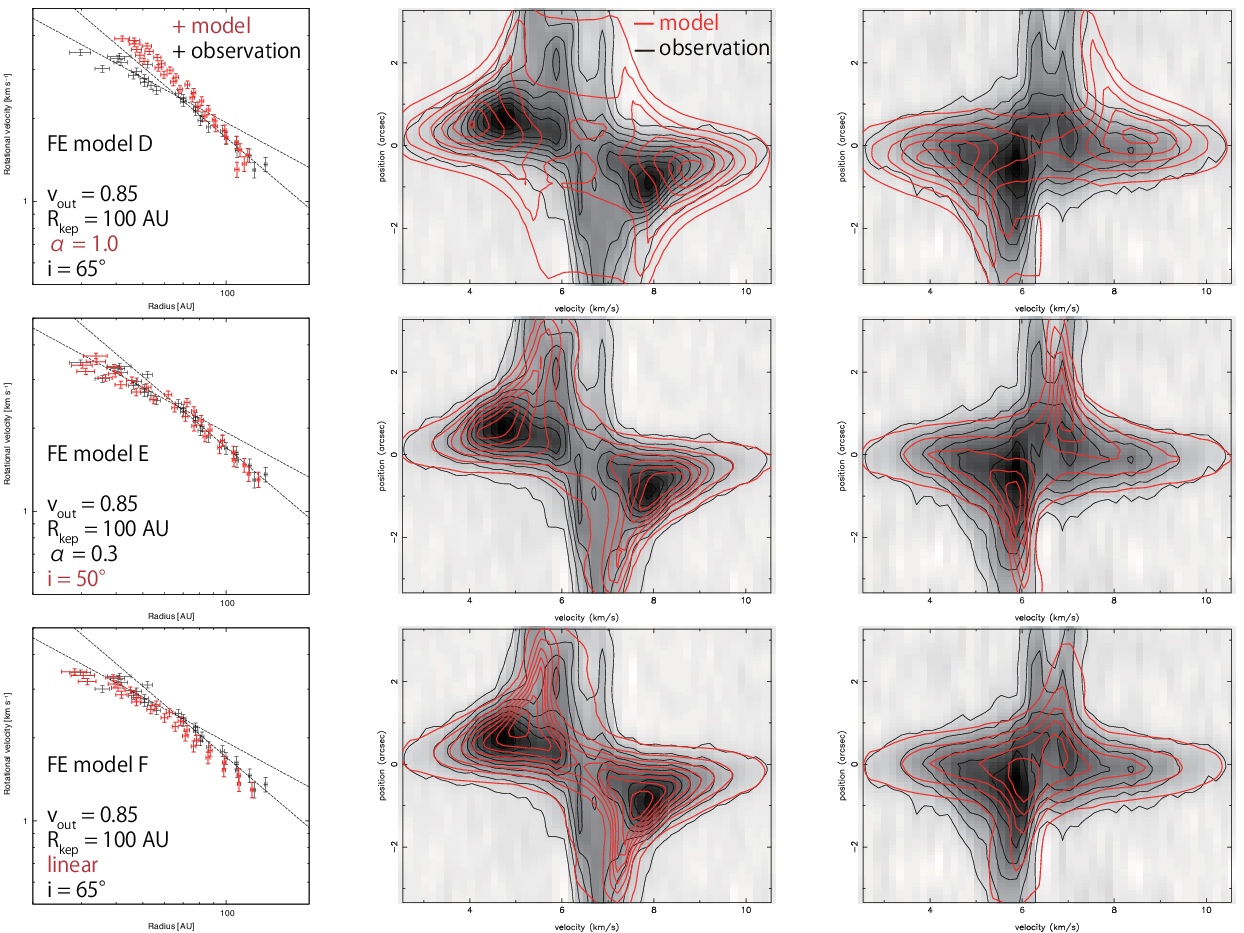}
\caption{(Continued)}
\end{figure}

\begin{figure}
\epsscale{1.0}
\plotone{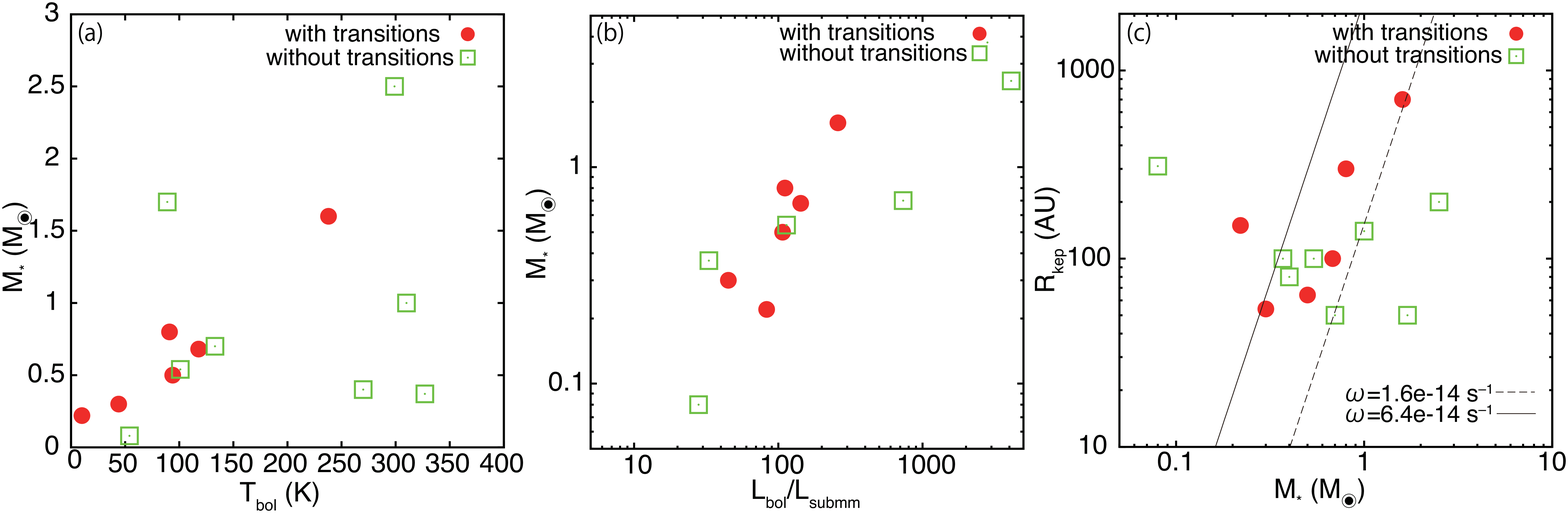}
\caption{Correlation between (a) bolometric temperatures and central protostellar masses, (b) $L_{bol}/L_{\rm submm}$ and central protostellar masses, and (c) central stellar masses and sizes of Keplerian disks around protostars. The data values are listed in Table \ref{tb:3}. Red filled circles corresponds to the protostars for which the transitions from infall motions to Keplerian rotations are identified well.\label{fig:rkms}}
\end{figure}

\begin{figure}
\epsscale{1.0}
\plotone{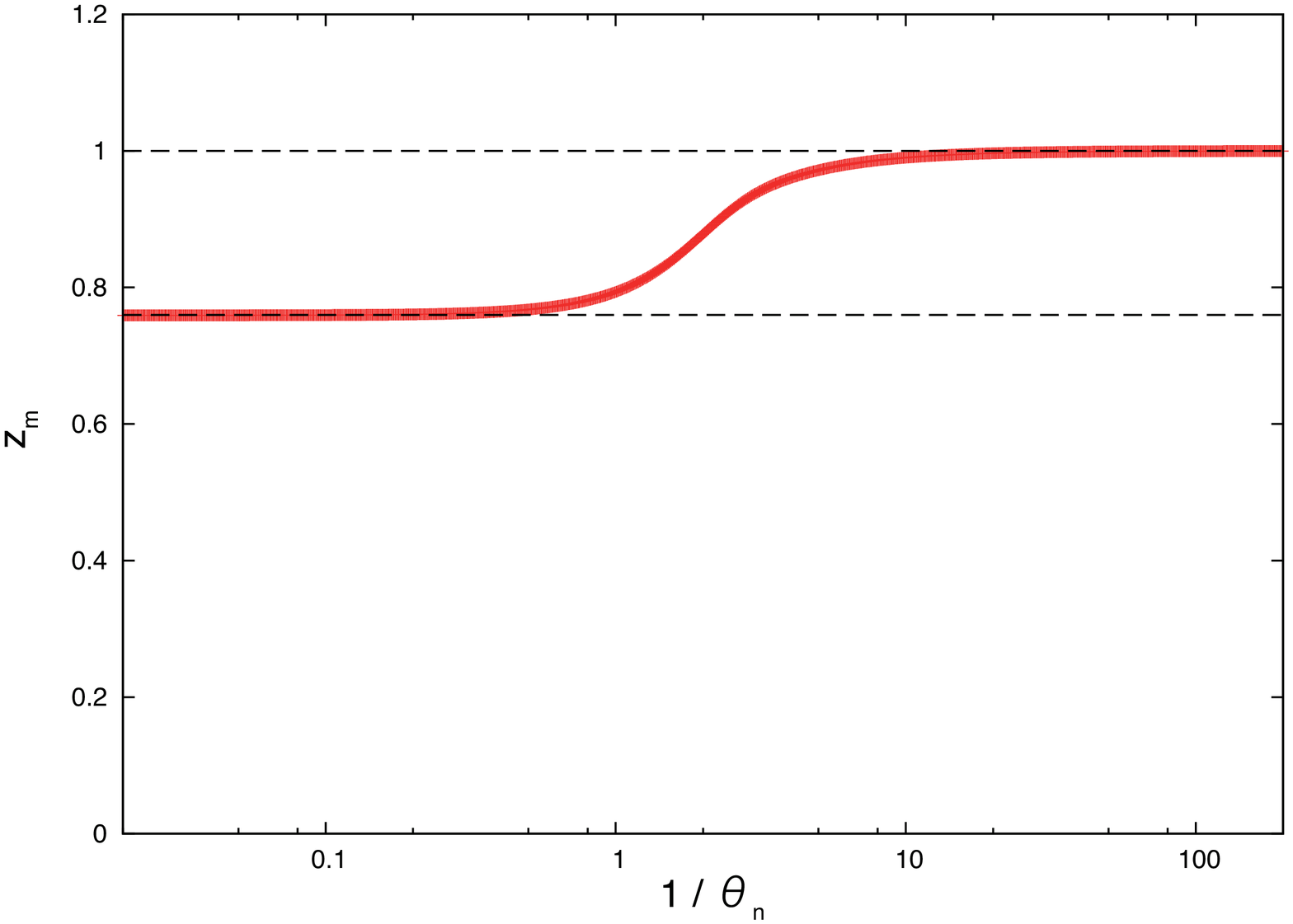}
\caption{The dependency of the normalized mean position $z_{m}=x_{m}/R_{o}$ on the normalized beam width $1/\theta _{n}=R_{o}\cos i/\theta $ in the case with $v=1/2$ (Keplerian rotation). \label{fig:z0}}
\end{figure}

\begin{figure}
\epsscale{1.0}
\plotone{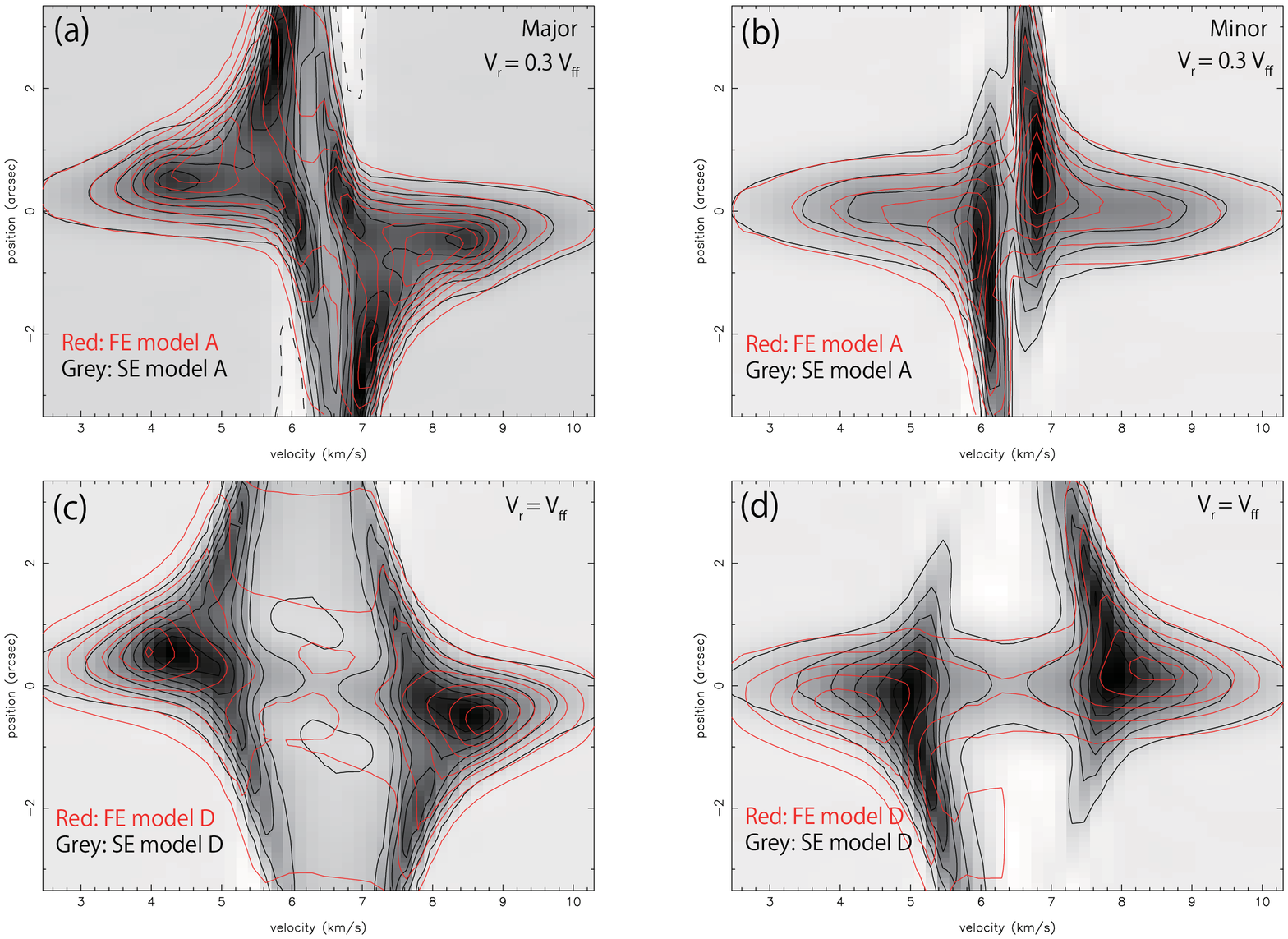}
\caption{Comparison FE model A/D in Sec. \ref{sc:inf} (red contours) and SE model A/D in Appendix \ref{sc:app3} (black contours and grayscales). The right and left columns show PV diagrams along the major and minor axes, respectively. The contour levels are the same as Fig. \ref{fig:pv} and Fig. \ref{fig:modinf}. \label{fig:saigo}}
\end{figure}

\clearpage
\begin{deluxetable}{cccc}
\tabletypesize{\scriptsize}
\tablecaption{Summary of the observational parameters\label{tb:1}}
\tablewidth{0pt}
\tablehead{\colhead{} & \colhead{CO $J=2-1$} & \colhead{C$^{18}$O $J=2-1$} & \colhead{continuum} \\
\colhead{Interferometer and date} & \multicolumn{3}{c}{ALMA (extended),  2012.Nov.6}}
\startdata
Target & \multicolumn{3}{c}{TMC-1A}\\
Coordinate center & \multicolumn{3}{c}{R.A. (J2000)=4$^{\rm h}$39$^{\rm m}$35$^{\rm s}\!\!$.010}\\
 & \multicolumn{3}{c}{Dec. (J2000)=25$^{\circ }41\arcmin 45\farcs 500$}\\
Frequency &230.5380 GHz&219.5604 GHz&225.4336 GHz \\
Primary beam & $27\farcs 3$&$28\farcs 6$&$27\farcs 9$ \\
Projected baseline length & \multicolumn{3}{c}{15.9 -- 285.7 m}\\
Synthesized beam (P.A.) & $1\farcs 02\times 0\farcs 90\ (-178^{\circ })$&$1\farcs 06\times 0\farcs 90\ (-176^{\circ })$&$1\farcs 01\times 0\farcs 87\ (+0.87^{\circ })$\\
Velocity resolution & 0.16 km$\,$s$^{-1}$ & 0.17 km$\,$s$^{-1}$ & 923 MHz\\
Noise level (no emission)& 5.6 mJy$\,$beam$^{-1}$ & 4.6 mJy$\,$beam$^{-1}$& ---\\
Noise level (detected channel)& 20 mJy$\,$beam$^{-1}$ & 7.1 mJy$\,$beam$^{-1}$& 0.96 mJy$\,$beam$^{-1}$\\
Passband calibrator & \multicolumn{3}{c}{J0522-364}\\
Flux calibrator & \multicolumn{3}{c}{Callisto}\\
Gain calibrator & \multicolumn{3}{c}{J0510+180}\\
\enddata
\end{deluxetable}

\begin{deluxetable}{cccccccc}
\tabletypesize{\scriptsize}
\tablecaption{Fixed and free parameters of the model fitting\label{tb:2}}
\tablewidth{0pt}
\tablehead{\colhead{fixed} & \colhead{$i$} &  \colhead{$R_{\rm in}$} & \colhead{$R_{\rm out}$} & \colhead{} & \colhead{} & \colhead{} & \colhead{}\\
\colhead{} & \colhead{$65^{\circ}$} & \colhead{0.1 AU} & \colhead{200 AU} & \colhead{} & \colhead{} & \colhead{} & \colhead{}}
\startdata
free & $M_{*p}$\tablenotemark{a} & $M_{\rm 200}$\tablenotemark{b} & $p$ & $T_{100}$ & $q$ & $R_{\rm cent}$ & $v$ \\
best & $0.56^{+0.05}_{-0.05}\ \Ms$ & $3.6^{+1.5}_{-1.0}\times 10^{-3}\ \Ms$ & $1.46^{+0.3}_{-1.0}$ & $38^{+6}_{-5}$ K & $-0.02^{+0.04}_{-0.04}$ & $166^{+400}_{-120}$ AU & $0.53^{+0.05}_{-0.05}$\\
\enddata
\tablenotetext{a}{$M_{*p}=M_{*}\sin ^{2}i$, not inclination corrected.}
\tablenotetext{b}{The mass within $R_{\rm out}=200$ AU. The mass of the Keplerian disk ($R_{\rm kep}=100$ AU) is calculated to be $2.5\times 10^{-3}\ \Ms$.}
\end{deluxetable}

\begin{landscape}
\begin{deluxetable}{cccccccccc}
\tabletypesize{\scriptsize}
\tablecaption{Parameters of Keplerian disks around protostars\label{tb:3}}
\tablewidth{0pt}
\tablehead{\colhead{Source}&\colhead{$L_{bol}\ (L_{\odot})$}&\colhead{$T_{bol}\ ({\rm K})$}&\colhead{$L_{bol}/L_{\rm submm}$\tablenotemark{a}}&\colhead{$R_{\rm kep}\ ({\rm AU})$\tablenotemark{b}}&\colhead{$M_{*}\ (M_{\odot})$\tablenotemark{c}}&\colhead{$M_{\rm disk}\ (M_{\odot})$\tablenotemark{d}}&\colhead{Class}&\colhead{transition?\tablenotemark{e}}&\colhead{References\tablenotemark{f}}}
\startdata
NGC1333 IRAS4A2 & 1.9\tablenotemark{g} & 51\tablenotemark{g} & 28\tablenotemark{g} & 310 & 0.08 &...& 0 & no & 1\\
VLA1623A & 1.1 & 10 & 83 & 150\tablenotemark{h} & 0.22 &...& 0 & yes & 2,13\\
L1527 IRS & 1.97 & 44 & 45 & 54 & 0.30 &0.0028-0.013& 0 & yes & 3,12,16 \\
R CrA IRS7B & 4.6 & 89 &...& 50 & 1.7 &0.024& I & no & 4 \\
L1551 NE & 4.2 & 91 & 111 & 300 & 0.8 &0.026& I & yes & 5,14,17 \\
L1551 IRS 5 & 22.1 & 94 & 107 & 64 & 0.5 &0.07& I & yes & 6,12,16 \\
TMC1 & 0.9 & 101 & 114 & 100\tablenotemark{i} & 0.54 &0.025-0.06& I & no & 7,12,16 \\
TMC-1A & 2.7 & 118 & 143 & 100 & 0.68 &$2.5\times 10^{-3}$& I & yes & 8,12,16 \\
TMR1 & 2.6 & 140 & 734 & 50 & 0.7 &0.01-0.015& I & no & 7,16 \\
L1489 IRS & 3.7 & 238 & 259 & 700 & 1.6 &3-7$\times 10^{-3}$& I & yes & 9,15,16 \\
L1536 & 0.4 & 270 &...& 80 & 0.4 &0.07-0.024& I & no & 7,12 \\
Elias 29 & 14.1 & 299 & 4215 & 200 & 2.5 &$\lessim 7\times 10^{-3}$& I & no & 10,12,16 \\
IRS 43 & 6.0 & 310 &...& 140 & 1.0 &$8.1\times 10^{-3}$& I & no & 11,15 \\
IRS 63 & 1.0 & 327 & 33 & 100 & 0.37 &0.055& I & no & 10,12,16 \\
\enddata
\tablenotetext{a}{$L_{\rm submm}$ is defined as the luminosity measured at wavelengths longer than 350 $\mu$m.}
\tablenotetext{b}{Outer radius of the Keplerian disk.}
\tablenotetext{c}{Central protostellar mass.}
\tablenotetext{d}{Mass of the Keplerian disk.}
\tablenotetext{e}{Frag showing whether or not sources show transitions from infall motions to Keplerian rotations. ``yes'' and ``no'' indicate sources with and without transitions, respectively.}
\tablenotetext{f}{References: (1) \citeauthor{cho2010} \citeyear{cho2010}; (2) \citeauthor{mu2013} \citeyear{mu2013}; (3) \citeauthor{oh2014} \citeyear{oh2014}; (4) \citeauthor{lin2014} \citeyear{lin2014}; (5) \citeauthor{ta2014} \citeyear{ta2014}; (6) \citeauthor{ch2014} \citeyear{ch2014}; (7) \citeauthor{ha2014} \citeyear{ha2014}; (8) this work; (9) \citeauthor{ye2014} \citeyear{ye2014}; (10) \citeauthor{lo2008} \citeyear{lo2008}; (11) \citeauthor{ei2012} \citeyear{ei2012}; (12) \citeauthor{kr2012} \citeyear{kr2012}; (13) \citeauthor{mu.la2013} \citeyear{mu.la2013}; (14) \citeauthor{fr2005} \citeyear{fr2005}; (15) \citeauthor{jo2009} \citeyear{jo2009}; (16) \citeauthor{gr2013} \citeyear{gr2013}; (17) \citeauthor{fr2005} \citeyear{fr2005}}
\tablenotetext{g}{Both $L_{bol}$ and $L_{\rm sub}$ are assumed to be a half of those derived toward IRAS 4A binary system. $T_{bol}$ is also assumed to be the same as that of the binary system.}
\tablenotetext{h}{\citet{mu2013} concluded that the Keplerian radius of VLA1623A is 150 AU though their UV-space PV diagram shows a turn over point of the rotational velocity profile at 50 AU. They argued that $R_{\rm kep}$ can be larger than 50 AU because of optical depth and absorption.}
\tablenotetext{i}{The analysis using $\log R-\log V$ diagram gives us a breaking radius of $\sim 90$ AU. The results here is, however, not affected significantly even if 90 AU is adopted instead of 100 AU.}
\end{deluxetable}
\end{landscape}
\end{document}